\documentclass[a4paper,11pt]{article}
\pdfoutput=1 
\usepackage{jheppub, bm, color} 
\usepackage[T1]{fontenc}
\usepackage{amssymb,amsfonts,slashed,amsthm,amsmath,graphicx, soul}
\title{Initial conditions for slow-roll inflation \\
in a random Gaussian landscape}

\author[1]{Ali Masoumi,}
\author[1]{Alexander Vilenkin,}
\author[1,2]{Masaki Yamada}

\affiliation[1]{Institute of Cosmology, Department of Physics and Astronomy, 
Tufts University, Medford, MA  02155, USA}

\affiliation[2]{Department of Physics, Tohoku University, 
Sendai, Miyagi 980-8578, Japan}

\emailAdd{ali@cosmos.phy.tufts.edu}
\emailAdd{vilenkin@cosmos.phy.tufts.edu}
\emailAdd{Masaki.Yamada@tufts.edu}

\def\({\left(}
\def\){\right)}
\def\[{\left[}
\def\]{\right]}

\def\pot{U}

\def\lmk{\left(}
\def\rmk{\right)}
\def\lkk{\left[}
\def\rkk{\right]}
\def\dd{{\rm d}}
\def\la{\left<}
\def\ra{\right>}
\newcommand{\eq}[1]{Eq.~(\ref{#1})}

\newcommand{\beq}{\begin{eqnarray}} 
\newcommand{\eeq}{\end{eqnarray}}

\newcommand{\bel}[1] {\begin{equation}\label{#1}}
\newcommand{\beal}[1] {\begin{eqnarray}\label{#1}}
\newcommand{\be}{\begin{equation}}
\newcommand{\ee}{\end{equation}}
\newcommand{\bea}{\begin{eqnarray}} 
\newcommand{\eea}{\end{eqnarray}}
\newcommand{\abs}[1]{\lvert#1\rvert}

\abstract{ In the landscape perspective, our Universe begins with a quantum tunneling from an
  eternally-inflating parent vacuum, followed by a period of slow-roll inflation. We investigate the
  tunneling process and calculate the probability distribution for the initial conditions and for the
  number of e-folds of slow-roll inflation, modeling the landscape by a small-field one-dimensional
  random Gaussian potential.  We find that such a landscape is fully consistent with observations, but
  the probability for future detection of spatial curvature is rather low, $P \sim 10^{-3}$.  }

\begin{document}
\maketitle
\flushbottom

\section{Introduction} 
\label{sect:Intro}

String theory predicts the existence of a vast landscape of vacuum states with diverse properties
\cite{Susskind:2003kw,BoussoPolchinski}.  In the cosmological context this leads to the picture of an
eternally inflating multiverse, where different spacetime regions are occupied by different vacua.
Transitions between the vacua occur through quantum tunneling, with bubbles of daughter vacuum nucleating
and expanding in the parent vacuum background.  According to this picture, our local region originated as
a result of tunneling from some inflating parent vacuum and then went through a period of slow-roll
inflation.  The number of vacua in the landscape is expected to be enormous, so predictions in this kind
of model must necessarily be statistical.  The properties of string theory landscape are not well
understood, and the approach adopted in much of the recent work is to substitute it by a scalar field
model with a random Gaussian potential
\cite{Tegmark,Easther,Frazer,Battefeld,McAllister:2012am,Yang:2012jf,Pedro:2013nda,Marsh:2013qca,Bachlechner,Wang,MV,Freivogel:2016kxc,Pedro:2016sli,EastherGuthMasoumi}.

In a recent paper \cite{MVY} we developed analytic and numerical techniques for studying the statistics
of slow-roll inflation in random Gaussian landscapes.  We applied these techniques to the simplest case
of small-field inflation in a one-dimensional random landscape.  In this case, inflation typically occurs
at local maxima or at inflection points of the potential \cite{LindeWestphal}.  Focusing mostly on the
inflection points, we found the probability distributions for the maximal number of inflationary e-folds
$N_{\rm max}$ and for the spectral index of density fluctuations $n_s$.

The maximal e-fold number $N_{\rm max}$ depends only on the shape of the potential near the inflection
point, but the actual number of e-folds, $N_e$, is sensitive to the initial conditions -- that is, to the
initial value $\phi_0$ of the inflaton field right after it tunnels from the parent vacuum.  If $\phi_0$
is too far away from the inflection point, the field may develop a large velocity and overshoot or it may
miss the slow-roll region entirely.  In the present paper we shall use numerical simulations to determine
the probability distribution for $\phi_0$ and to investigate its effect on the statistical properties of
inflation.  As before, we shall restrict our analysis to the simplest case of one-dimensional potentials.

In the next Section we review some general properties of random Gaussian potentials, and in Sec.~3 we summarize earlier work on inflection-point inflation.  Our numerical simulations and the results for the distribution of $\phi_0$ are presented in Sec.~\ref{sec:initial condition}.  In Sec.~\ref{sec:1D} we develop a semi-analytic method to study the evolution of the scalar field after tunneling.  We find the probability distribution for the number of e-folds of slow-roll inflation $N_e$ and discuss the implications of our results for the prospects of detection of spatial curvature.  Our conclusions are summarized and discussed in Sec.~\ref{sec:Conclusion}.
Some technical details related to the simulations are relegated to the appendices.  Throughout the paper we use reduced Planck units with $8\pi G=1$.

\section{Random Gaussian landscapes}\label{sec:Random}

Consider a one-dimensional random Gaussian landscape model with 
a potential $U(\phi)$ satisfying the following correlation function: 
\bel{Correlation}
	\langle \pot (\phi_1) \pot(\phi_2)\rangle=  F (|\phi_1 - \phi_2|)=\frac1{2\pi}\int \dd {k} \,P(k) e^{i{k} (\phi_1-\phi_2)}~. 
\ee
We specifically consider a Gaussian-type correlation function defined as 
\be
F(\phi)=U_0^2 e^{-\phi^2/2\Lambda^2},
\label{correlation function}
\ee
with $\Lambda$ playing the role of the correlation length in the landscape.  
Then the spectral function $P(k)$ is given by 
\be
\label{pk}
P(k)= U_0^2 (2\pi\Lambda^2)^{1/2} e^{-\Lambda^2 k^2/2} .
\ee
We define different moments of the spectral function as  
\beq
\label{sigmaDef}
	\sigma_{n}^2 = \frac1{2\pi}\int d k (k^2)^n P(k)= \frac{2^n \Gamma \lmk n + \frac{1}{2} \rmk }{\Gamma \lmk \frac{1}{2} \rmk} 
	\frac{U_0^2}{\Lambda^{2n}}, 
\eeq
where we used \eq{pk} in the second line. 

Once we specify a set of points $\{\phi_1, \ldots, \phi_n\}$ and define $U_j \equiv U(\phi_j)$, 
the probability distribution for $U_j$ is given by 
\bel{dist}
	P(U) = \frac{\sqrt{{\rm det} M^{-1}_{ij}}}{(2\pi)^{n/2}} e^{-\frac12 U_i M^{-1}_{ij}U_j}~, 
\ee
where the positive definite matrix $M$ is an $n\times n$ matrix of correlators defined by 
\bel{Correlations2}
M_{ij} \equiv F\( |\phi_i -\phi_j |\)~. 
\ee

Since $M_{ij}$ is a symmetric matrix, we can diagonalize it by an orthogonal matrix $\cal O$. 
Then the probability distribution for variables $\beta_i \equiv {\cal O}_{ij} U_j$ is given by 
\bel{dist2}
	P(\beta) = \frac{\sqrt{\prod_i \lambda_i }}{(2\pi)^{n/2}} e^{-\frac12 \lambda_i \beta_i^2}~, 
\ee
where $\lambda_i$'s are eigenvalues of the matrix $M_{ij}$. 
Such random variables can be easily generated in numerical simulations. 

When the values of $U(\phi_j)$ are generated at a sufficient density on the $\phi$-axis, we can
interpolate them to obtain a smooth potential.  We check in the App.~\ref{sec:singleValued} that this
procedure typically saturates at a few points per correlation length $\Lambda$: the interpolated
potential changes very little with the addition of more points.  We shall use this method with four
points per correlation length to generate realizations of a random Gaussian landscape.

\begin{figure}[t] 
   \centering
   \includegraphics[width=2.5in]{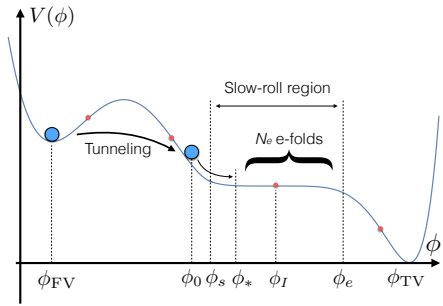} 
   \caption{
A schematic picture of the kind of potential we are interested in. 
Inflection points are marked by red dots. 
}
   \label{fig:schematic}
\end{figure}

Assuming that our universe is a result of a bubble nucleation event, we anticipate that the state of the
universe after nucleation allows for a prolonged period of inflation. This is possible for parts of the
landscape that resemble Fig.~\ref{fig:schematic}, which includes a parent vacuum and a daughter vacuum
separated by a barrier.  We shall refer to parent and daughter vacua as ``false'' (FV) and ``true'' (TV)
vacua, respectively.  The inflection points where $U''=0$ are marked by red dots in the figure.  In this
example, there are three inflection points between the top of the barrier and the true vacuum.  The
potential is rather flat near the middle inflection point ($U'$ and $U''$ are small), and slow-roll
inflation can be expected to occur in this region.

We are going to focus on the case when the correlation length of the potential is small compared to the
Planck scale, $\Lambda\ll 1$.  Inflation in this case is typically of the small-field type: the slow-roll
occurs in a narrow range $\Delta\phi\ll\Lambda$ and the potential during the slow-roll remains
approximately constant.\footnote{Large-field inflation requires a long stretch of flat potential with
  $\Delta\phi \gg 1$.  Such stretches will occur in a random landscape, but for $\Lambda\ll 1$ they will
  be very rare.}

The correlation function (\ref{correlation function}) implies that the average value of the potential is
$\langle U(\phi)\rangle=0$, so positive and negative values of $U(\phi)$ are equally likely.  In this
case, the minima of the potential are predominantly at negative values of $U$.  This effect is especially
pronounced for higher-dimensional random landscapes, where positive-energy vacua may not exist at all
\cite{BrayDean,Battefeld}.  This problem can be alleviated by adding a constant shift term to the
potential, $U(\phi)\to U(\phi)+C$.  One can also consider models where $C$ is variable, but its
characteristic scale of variation is much greater than $\Lambda$.  For example, in axionic landscapes
this term can have the form $C(\phi)=m^2\phi^2 /2$ with a very small mass $m$ \cite{MV} or
$C(\phi)\propto \cos(\mu\phi)$ with $\mu\ll 1/\Lambda$ \cite{evil}.  Then the local properties of the
potential are still determined by the correlator (\ref{correlation function}), while the extra term
provides an additive shift, which is characterized by its own probability distribution.

Here, we are interested in the case when the true vacuum has an almost vanishing vacuum energy.  In a
generic landscape, the number of such vacua is extremely small, and trying to find them by random
sampling is a hopeless task.  In our numerical simulations we simply added a constant to the generated
random potential, so that $U = 0$ in the true vacuum.  We expect the resulting ensemble of realizations
to be similar to what one would get by random sampling in a landscape with a flat distribution of the
shift parameter $C$.

\section{Analytic results for inflection-point inflation}
\label{sec:taylor}

 Inflection-point inflation was first studied by Baumann et al \cite{Baumann}.  Here we review some of their results, which will be useful in subsequent sections.


The necessary conditions for slow-roll inflation are $\epsilon_s, \eta_s\ll 1$, 
where 
\beq
 &&\epsilon_s = \frac{1}{2} \lmk \frac{U'}{U} \rmk^2,
 \label{epsilon}
 \\
 &&\eta_s = \frac{U''}{U}. 
\label{eta}
\eeq Note that the values of the slow-roll parameters at a randomly chosen point in the landscape are
typically given by $\epsilon_s \sim \eta_s \sim \Lambda^{-2} \gg 1$, so inflation can occur only in rare
regions of the landscape.

To study inflection-point inflation, we approximate the potential by a third-order Taylor expansion,
\beq
 U(\phi) = U_{\rm I} + {U_{\rm I}}' \phi + \frac{1}{3!} {U_{\rm I}}''' \phi^3, 
 \label{cubic}
\eeq where the subscript $I$ indicates the value at the inflection point $\phi_{\rm I}$, which is set to
be at $\phi_{\rm I}=0$: ${U_{\rm I}}''=0$.  The slow-roll conditions require that ${U_{\rm I}}'\ll U_{\rm
  I}$, but ${U_{\rm I}}'''$ does not need to be small, and we shall assume it to be comparable to the rms
value, ${U_{\rm I}}'''\sim U_0/\Lambda^3$.  The slow-roll region is then determined mainly by the
condition $\eta_s\ll 1$ and can be specified as $[\phi_s, \phi_e]$, where \beq \phi_s \sim - \frac{U_{\rm
    I}}{\abs{{U_{\rm I}}'''}},
 \label{sr1}
 \\
 \phi_e \sim \frac{U_{\rm I}}{\abs{{U_{\rm I}}'''}}. 
\label{sr}
\eeq We assume that ${U_{\rm I}}'{U_{\rm I}}''' >0$, since otherwise the potential has a shallow
high-energy minimum near the inflection point and inflation drives the field into that minimum
\cite{MVY}.

It follows from (\ref{sr1}),(\ref{sr}) that the size of the slow-roll region is typically $\Delta\phi\sim
\Lambda^3 \ll \Lambda$.  This justifies the assumption that $U(\phi)$ can be approximated by a constant
in this region.

The ``maximal e-folding number'' can be defined as 
\beq
 N_{\rm max} =
 - \int_{\phi_s}^{\phi_e} d\phi\frac{U(\phi)}{U'(\phi)} 
 \approx
 - \int_{-\infty}^\infty d\phi\frac{U(\phi)}{U'(\phi)} 
 \approx \pi \sqrt{2} \frac{U_{\rm I}}{\sqrt{{U_{\rm I}}'{U_{\rm I}}'''}}, 
 \label{Nmax}
\eeq
where we have assumed that $N_{\rm max} \gg 1$ (which is necessary for extending the integration to $\pm\infty$).

The spectral index $n_s$ can be expressed in terms of $N_{\rm max}$ as 
\beq
 &&n_s \simeq 1 - \frac{4\pi}{N_{\rm max}} {\rm cot} \lmk \frac{\pi N_e^{(\rm CMB)}}{N_{\rm max}} \rmk, 
 \label{n_s for inflection}
\eeq where $N_e^{(\rm CMB)}$ ($\simeq 50$-$60$) is the e-folding number at which the CMB scale leaves the
horizon.  It follows from (\ref{n_s for inflection}) that $n_s$ is greater than $1-4/N_e^{(\rm CMB)}
\approx 0.92$, which is approached in the limit $N_{\rm max}\to\infty$.  Note that for hilltop inflation
$n_s <0.92$ \cite{MVY}, while the observed value is $n_s\approx 0.97$ and lies in the inflection-point
range.  This provides additional motivation to focus our analysis on inflection-point inflation.

The magnitude of density fluctuations is given by
\beq
\Delta_R^2=\frac{1}{12\pi}\frac{U_{\rm I}^3}{{{U_{\rm I}}'}^2} =\frac{N_{\rm max}^4}{48\pi^6}\frac{{{U_{\rm I}}'''}^2}{U_{\rm I}} \sim \frac{U_0 N_{\rm max}^4}{48\pi^6 \Lambda^6},
\eeq
where in the second step we used Eq.~(\ref{Nmax}) for $N_{\rm max}$ and in the last step we assumed that $U_{\rm I}$ and ${U_{\rm I}}'''$ have their typical values.  The observed value of $\Delta_R^2 \sim 4\times 10^{-9}$ can be obtained by adjusting the parameters $U_0$ and $\Lambda$:  
\beq
\frac{U_0}{\Lambda^6} \sim 10^{-12}\left(\frac{100}{N_{\rm max}}\right)^4 \lesssim 10^{-12}.
\label{ULambda}
\eeq

The probability distributions for $N_{\rm max}$ and $n_s$ in a random Gaussian landscape have been calculated in Ref.~\cite{MVY}.  Here we quote the results: 
\beq
 &&P(N_{\rm max}) \propto N_{\rm max}^{-3},
 \label{PNmax}
 \\
 &&P (n_s) = P(N_{\rm max}) \frac{\dd N_{\rm max}}{\dd n_s}.
 \label{Pns}
\eeq

The distribution (\ref{PNmax}) gives the probability that a randomly chosen inflection point in the
landscape is characterized by a given value of $N_{\rm max}$.  Note, however, that here we are interested
only in inflection points located between a high-energy false vacuum and a zero-energy true vacuum, as
shown in Fig.~\ref{fig:schematic}.  As explained in Sec.~\ref{sec:Random}, we obtain such configurations
by adding a constant term to a randomly generated potential.  This procedure changes the value of $U$ at
the inflection point; hence it affects the value of $N_{\rm max}$ in Eq.~(\ref{Nmax}) and may potentially
affect the distribution (\ref{PNmax}).  We shall see, however, that the form of this distribution remains
unchanged.

As we mentioned in the Introduction, the actual number of inflationary e-folds $N_e$ depends on the
initial conditions after tunneling and is generally different from $N_{\rm max}$.  We shall find the
probability distributions for the initial conditions and for $N_e$ in the following sections.

\section{Initial conditions for inflation}
\label{sec:initial condition}

\subsection{General formalism}
\label{sec:general formalism}

Decay of the false vacuum occurs through bubble nucleation, which is a quantum tunneling process.  In the
semiclassical approximation, the tunneling is described by an $O(4)$-symmetric instanton $\phi(r)$, which
can be found by solving the Euclidean field equation \cite{Coleman} \bel{quarticPot}
\frac{d^2\phi}{dr^2}+ \frac3r \frac{d\phi}{dr}=\frac{dU}{d\phi}~.  \ee Here we assume that gravitational
effects on the tunneling can be neglected, which is usually the case in a small-field
landscape.\footnote{As a rule of thumb, gravitational effects are unimportant when the nucleating bubble
  is much smaller than the Hubble radius.  The typical size of a bubble at nucleation is $R\sim
  |U''|^{-1/2}$, while the Hubble radius is $H^{-1}\sim U^{1/2}$, where all quantities are evaluated at
  the top of the barrier.  Requiring that $R\ll H^{-1}$, we have $U/|U''|\sim \Lambda^2 \ll 1$, which is
  satisfied for small field inflation.}  The boundary conditions for $\phi(r)$ are $\phi'(r = 0) = 0$ and
$\phi(r\to\infty)=\phi_{\rm FV}$, where $\phi_{\rm FV}$ is the value of $\phi$ in the false vacuum.  The
tunneling probability is determined mostly by the exponential factor, \beq P_{\rm tunneling} \propto
e^{-S_E}, \eeq where $S_E$ is the Euclidean instanton action \beq S_E = \int_0^\infty \dd r 2 \pi^2 r^3
\lmk \frac{1}{2} \phi'^2 + U(\phi(r)) - U_{\rm FV} \rmk, \eeq and $U_{\rm FV} \equiv U(\phi_{\rm FV})$.

It will be convenient to introduce dimensionless variables ${\bar\phi}$ and ${\bar r}$ as
\beq
 &&\phi =  \Lambda {\bar\phi},
 \label{rescale of phi}
 \\
 &&r =  \Lambda U_0^{-1/2} {\bar r}, 
 \label{rescale of r}
 \\
 &&U(\phi)= U_0 {\bar U}({\bar\phi}).
 \label{rescale of U}
\eeq
In terms of the new variables, Eq.~(\ref{quarticPot}) still has the same form,
\bel{phibareq}
\frac{d^2{\bar\phi}}{d{\bar r}^2}+ \frac{3}{\bar r} \frac{d\bar\phi}{d{\bar r}}=\frac{d{\bar U}}{d{\bar\phi}}~,
\ee
where the potential ${\bar U}({\bar\phi})$ is now characterized by the correlation function (\ref{correlation function}) with $U_0 = \Lambda =1$.  We note that 
\beq
N_{\rm max}= \Lambda^2 {\bar N}_{\rm max} ,
\label{barNmax}
\eeq
where ${\bar N}_{\rm max}$ is the maximal e-folding number for the rescaled potential ${\bar U}({\bar\phi})$. 
We also define the action $\bar{S}$ for the rescaled variables: 
\beq
 S_E = \frac{\Lambda^4}{U_0} \bar{S}. 
\label{rescaleS}
\eeq

The initial value $\phi_0$ of the inflaton field after tunneling is set by the value of ${\bar\phi}$ at the center of the instanton, 
\beq
\phi_0 = \Lambda{\bar\phi}(0) .
\label{phi0}
\eeq
The probability distribution for ${\phi}_0$ can now be found with the aid of numerical simulations.

It is well known that instanton solutions for multi-dimensional field spaces are not unique, because
there may be more than one saddle point between the true vacuum and false vacuum.  As we explain in
App.~\ref{sec:uniqueness}, there may also be multiple instanton solutions in the one-dimensional case,
but for a different reason.  For a generic potential, there is typically a single instanton describing
tunneling to a close vicinity of the true vacuum.  In the presence of a flat inflection region,
additional instantons may appear, corresponding to tunneling to the neighborhood of the inflection point
$\phi_{\rm I}$.  As we make the inflection region flatter, at some point the instanton tunneling to the
true vacuum disappears, and only tunneling to the vicinity of inflection point remains possible.
  
\begin{figure}[t] 
   \centering
   \includegraphics[width=2.5in]{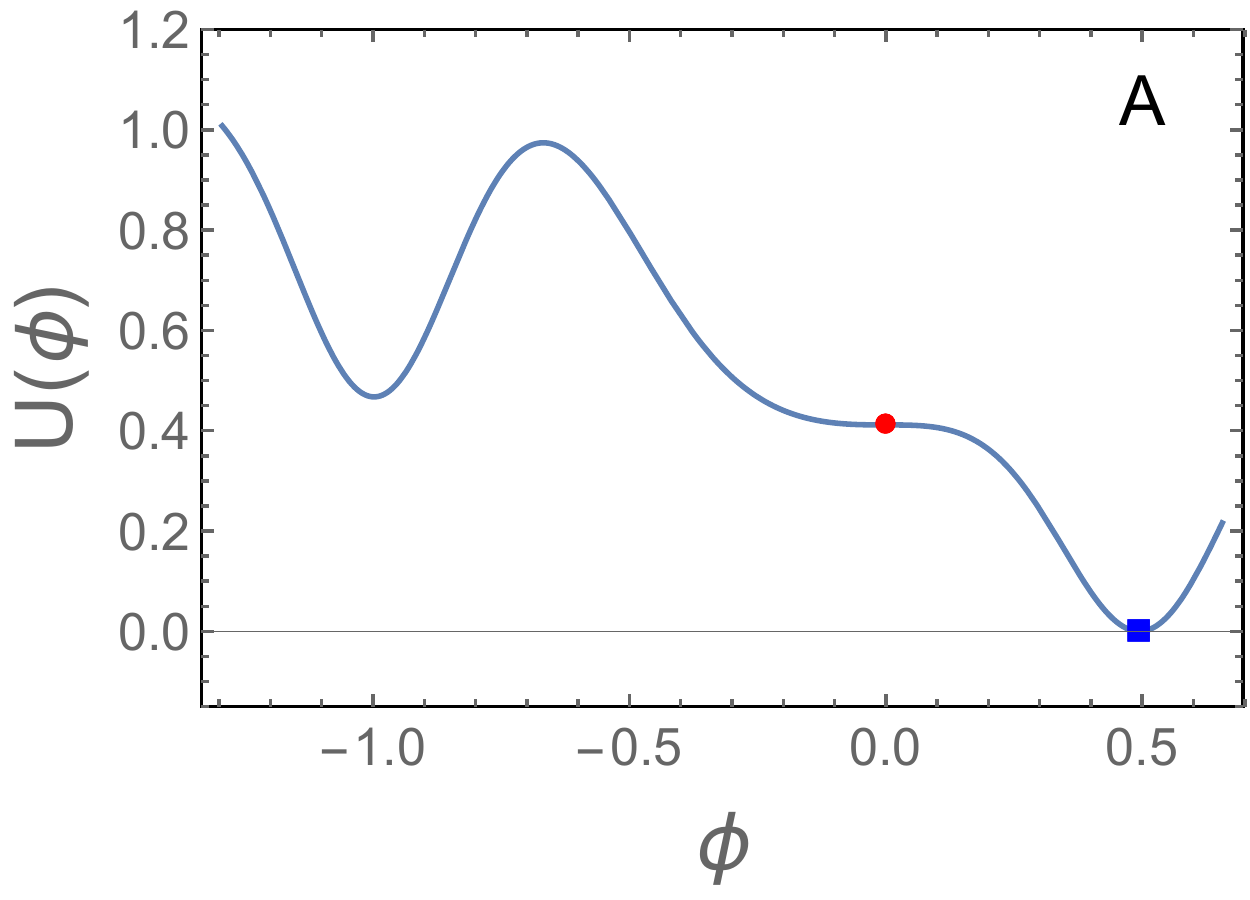} 
   \qquad 
   \includegraphics[width=2.5in]{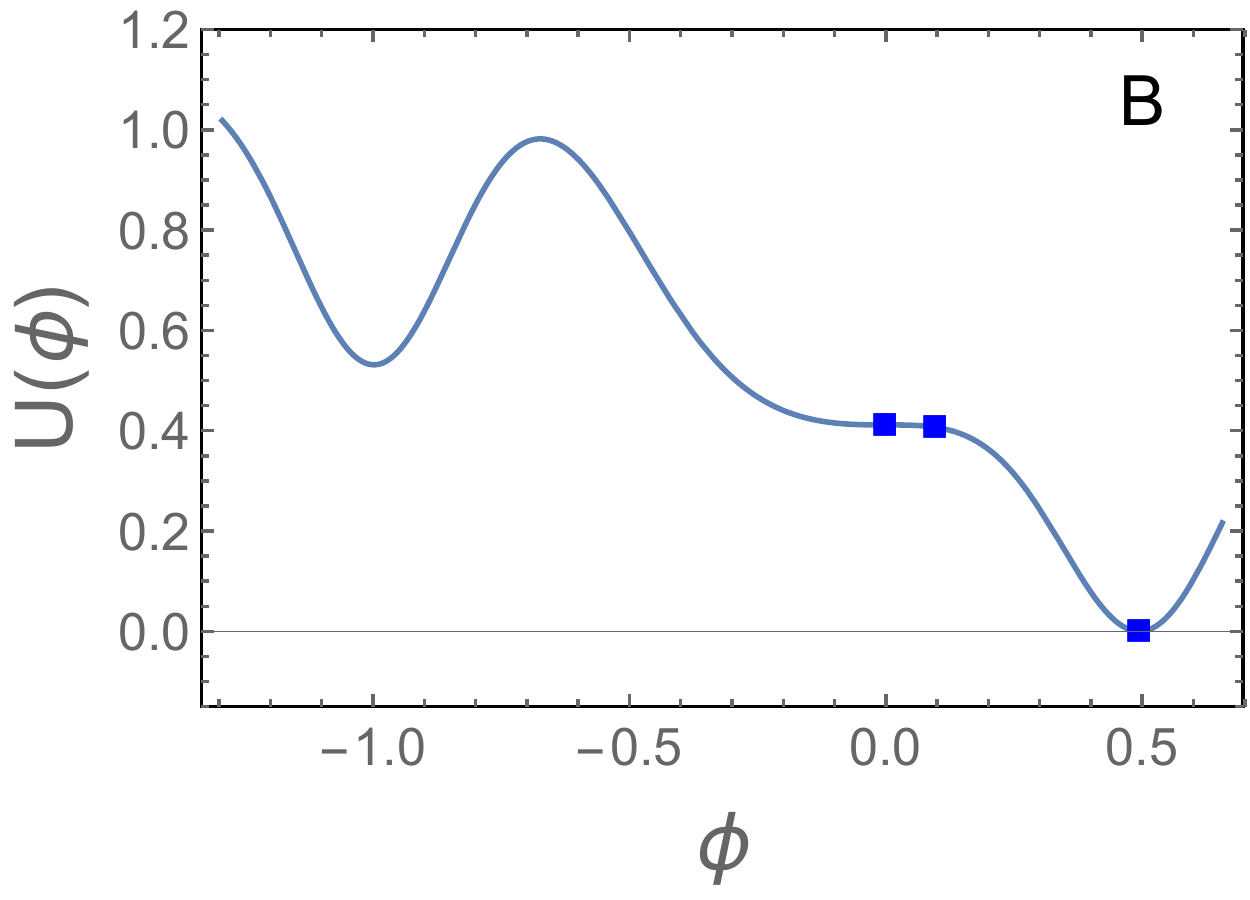}    
   \\
   \includegraphics[width=2.5in]{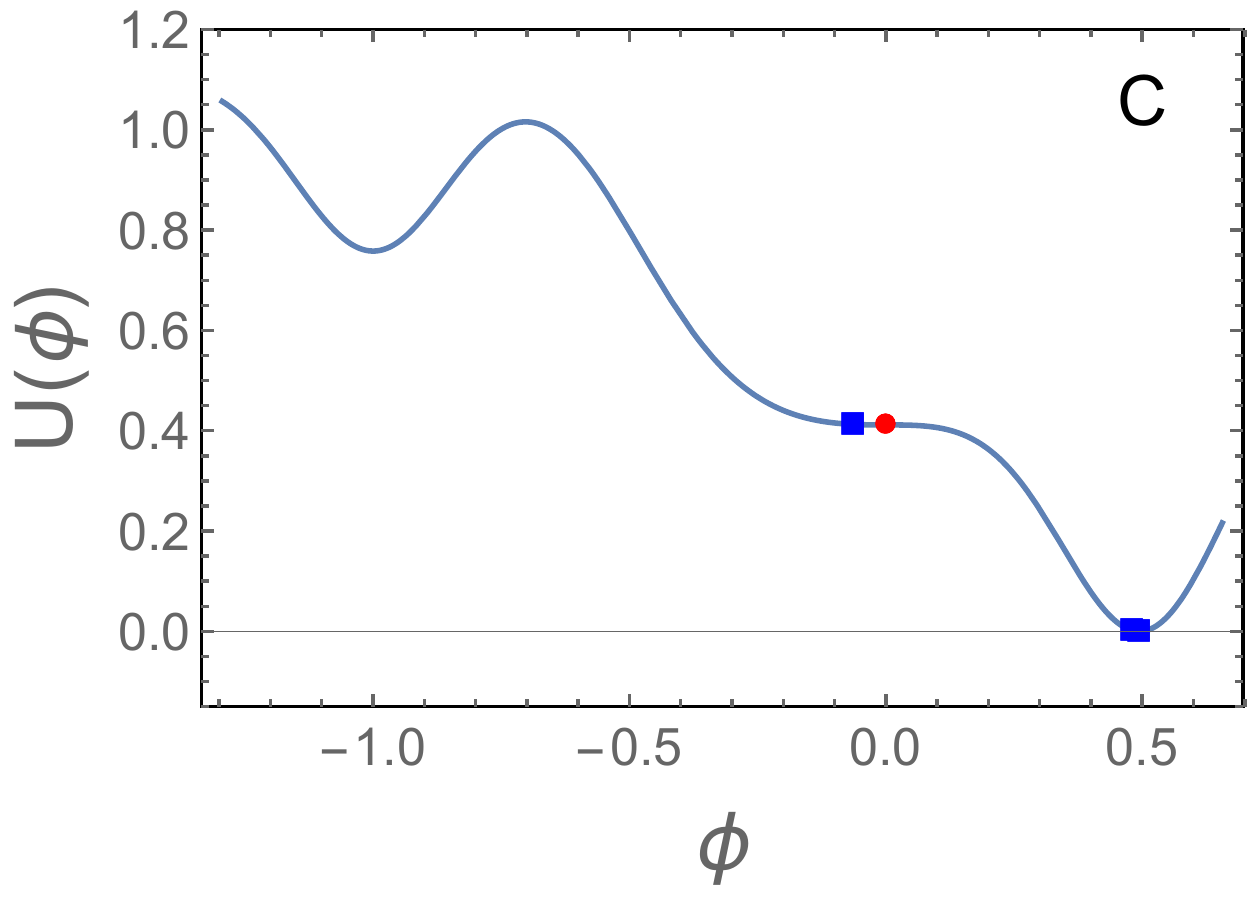} 
   \qquad 
   \includegraphics[width=2.5in]{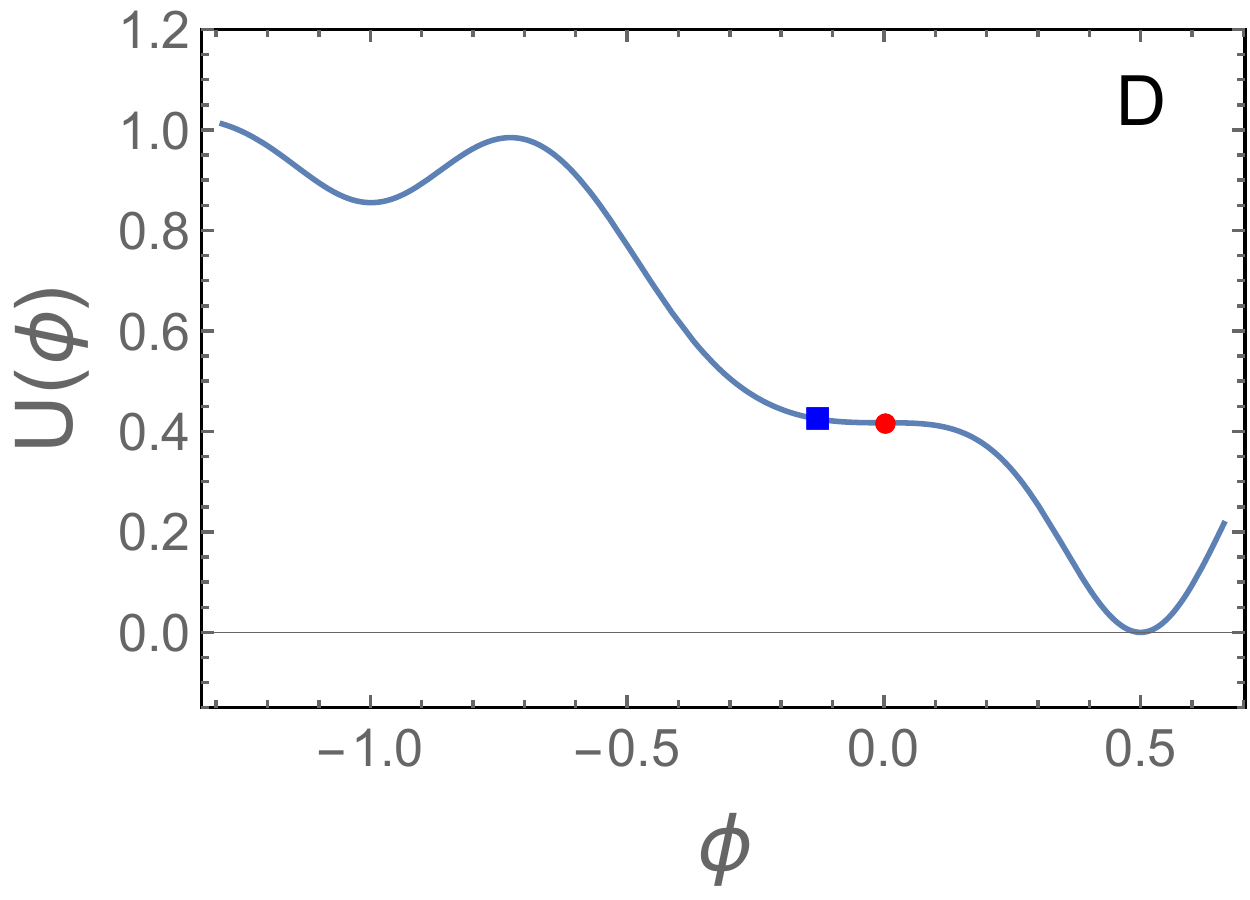}    
   \caption{
Tunneling points, represented by blue squares, are shown for potentials having similar shapes but 
different values of $U_{\rm FV}$.  The relevant inflection points are marked by red dots. 
The upper left figure has a single tunneling to the true vacuum, the upper right and lower left figures have tunneling points to neighborhoods of both $\phi_{\rm TV}$ and $\phi_{\rm I}$.  The bottom right potential has tunneling only to the neighborhood of $\phi_{\rm I}$.
}
   \label{fig:schematic2}
\end{figure}

The number and character of the instantons also depend on the shape and height of the potential barrier.
As an illustration we show some examples in Fig.~\ref{fig:schematic2}, where the tunneling points
$\phi_0$ are indicated by blue squares.  All four potentials in the figure are rather similar, except
they have different values of the false vacuum energy density $U_{\rm FV}$.  In the upper left frame,
$U_{\rm FV}$ is almost degenerate with $U_{\rm I}$ and there is a single instanton solution, which brings
$\phi$ almost all the way to the true vacuum.  In the upper right frame, $U_{\rm FV}$ is somewhat higher
and additional instantons appear, which describe tunneling with $\phi_0$ close to the inflection point.
As we explain in App.~\ref{sec:uniqueness}, additional instanton solutions appear in pairs.  As $U_{\rm
  FV}$ gets higher, the middle tunneling point moves towards the true vacuum tunneling point, and
eventually the two points ``annihilate''.  In the lower right frame, $U_{\rm FV}$ is still higher, and
tunneling is now possible only to the neighborhood of the inflection point.

When several instantons are present, one can compare the instanton actions to determine the dominant
decay channel.  We found that tunneling to the true vacuum dominates in most of these cases.\footnote{In
  Appendix.~\ref{sec:tunnelingFar} we show an example where it is favorable to tunnel to far away minima,
  even if there are local minima in between.}  We note, however, that in the present context we are
interested in tunnelings that lead to sufficiently long inflation, regardless of their relative rate
compared to other tunneling processes.

\begin{figure}[t] 
   \centering
   \includegraphics[width=2.5in]{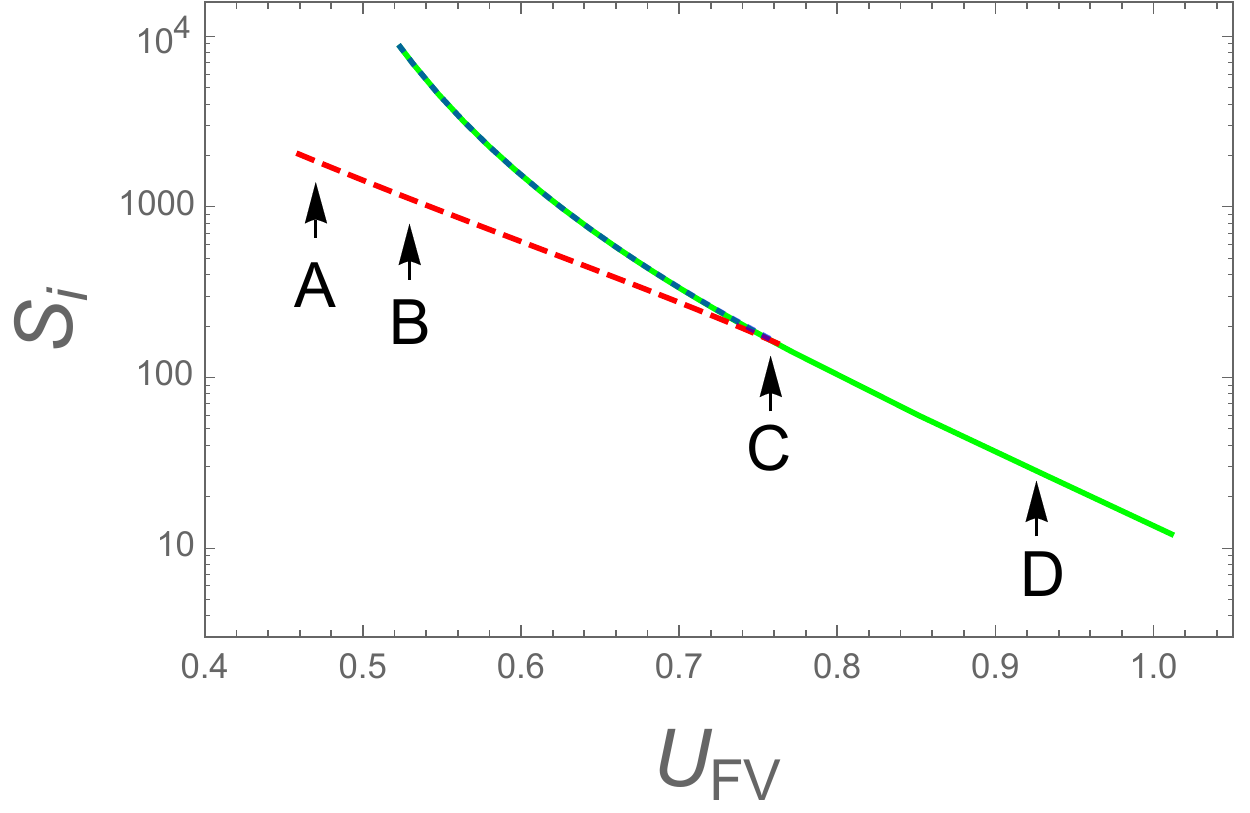} 
   \qquad
   \includegraphics[width=2.5in]{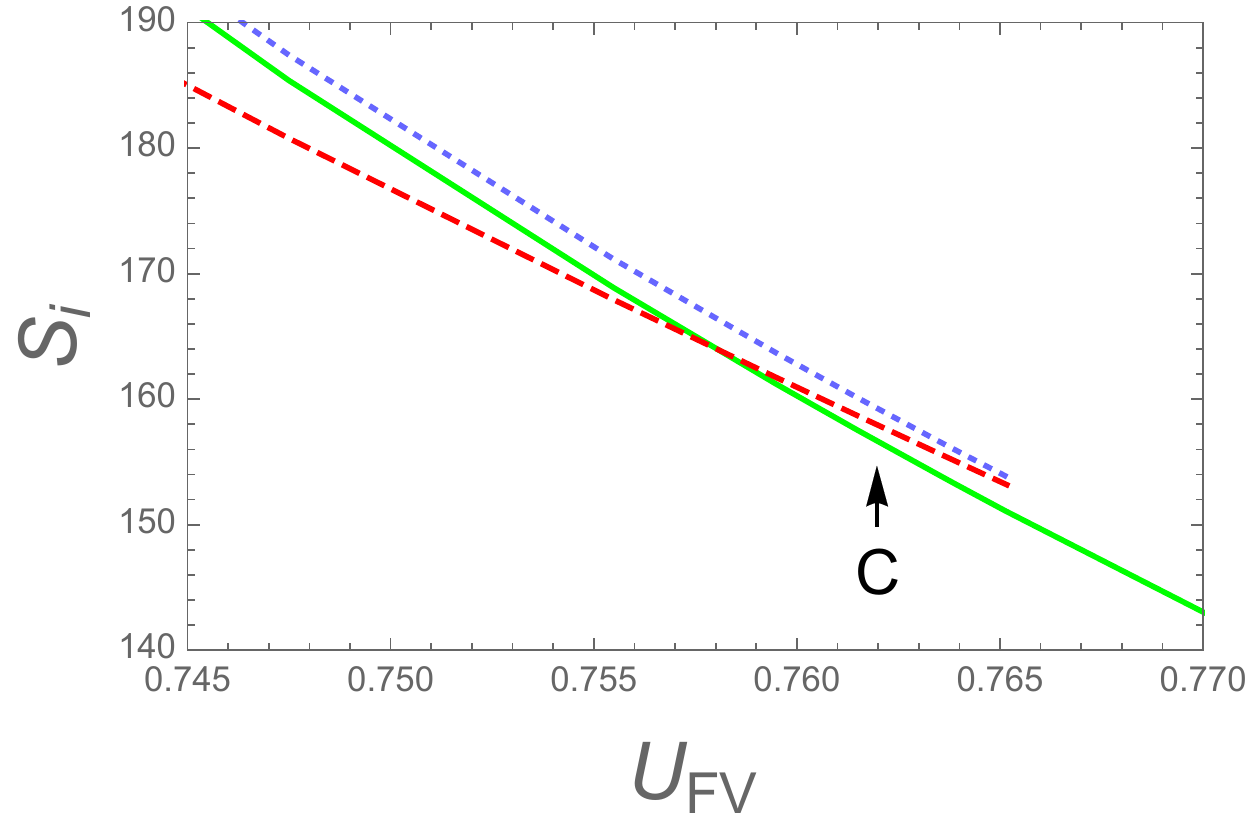} 
   \caption{
   Action of different instanton solutions as a function of $U_{\rm FV}$ with other features of potential fixed, as explained in the text. 
}
   \label{fig:S-V}
\end{figure}

To illustrate the dependence of the action of different instantons on the shape of the potential, we fix
the potential on the right side of the barrier in Fig.\ref{fig:schematic2} and calculate the instanton
actions for different values of $U_{\rm FV}$.  The result is shown in Fig.~\ref{fig:S-V}.  The red dashed
(green solid) line is the action of the instanton solution that brings $\phi$ close to the true vacuum
(inflection point). The blue dotted line is the one with $\phi_0$ between the other two tunneling points.
We see that tunneling to the true vacuum dominates in (almost) the entire range where the corresponding
instanton exists.  The blue dotted line is always just above the green solid line; they are so close that
they appear to coincide in the left frame of the figure.  When $U_{\rm FV}$ increases to a certain
threshold, the blue dotted line meets and annihilates with the red dashed line (see the right
frame). Before the annihilation, the red dashed line crosses the green solid line, which means that
tunneling to the inflection point becomes dominant. However, the region where this occurs is so small
that we cannot see it in the left frame.  Since the instanton solution with $\phi_0$ between the other
two tunneling points is always subdominant, we neglect it in the rest of the paper.  We denote the
instanton action for solutions that bring $\phi$ close to the true vacuum (inflection point) as $S_{\rm
  T}$ ($S_{\rm I}$).

\subsection{Numerical simulation}

We used the procedure outlined in Sec.~\ref{sec:Random} to generate $2.0 \times 10^9$ segments of a
random Gaussian landscape, with each segment having length $8\Lambda$.  The number of local minima in
such a segment is almost always greater than or equal to two. We set the number of points per correlation
length at $4$, which means that every segment contains 33 points $\phi_j$.  We generate the values of the
potential $U(\phi_j)$ at these points according to the probability distribution (\ref{dist}) and use a
fifth order spline to interpolate between them.

\begin{figure}[htbp]
  \centering
  \includegraphics[width=2.2in]{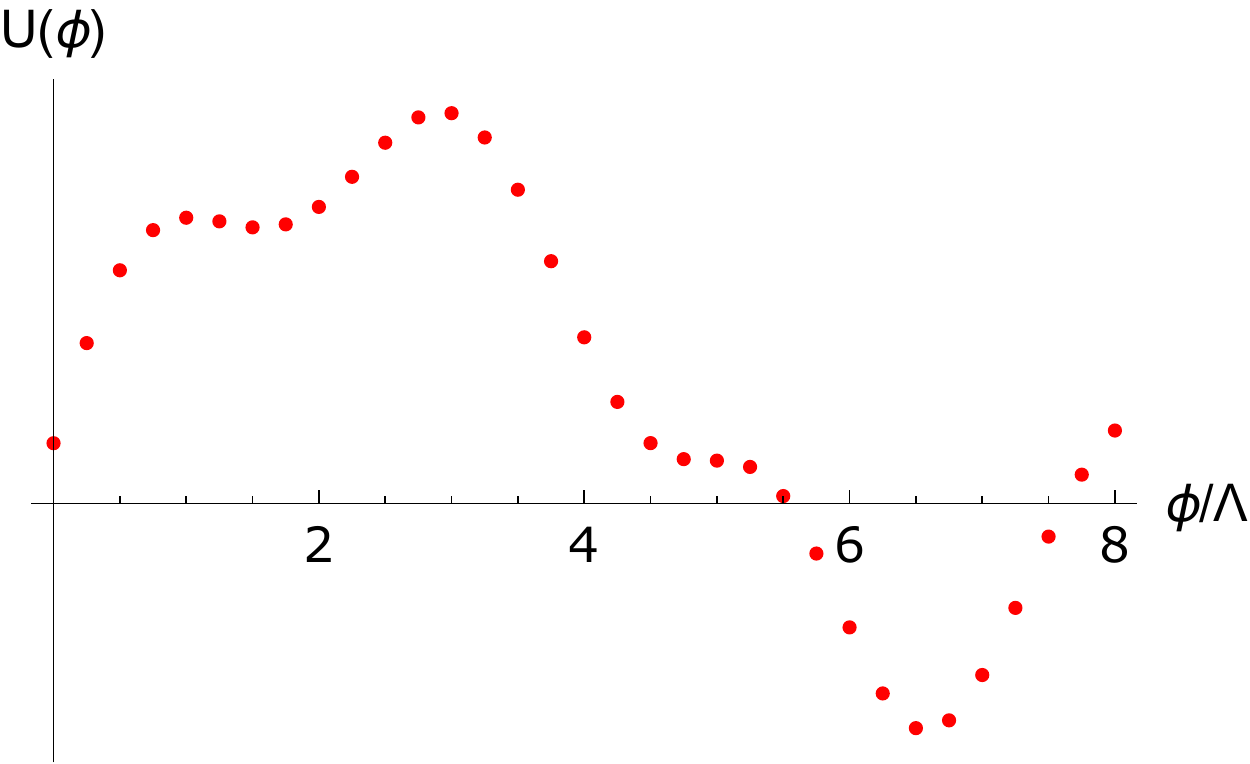}
  \includegraphics[width=2.2in]{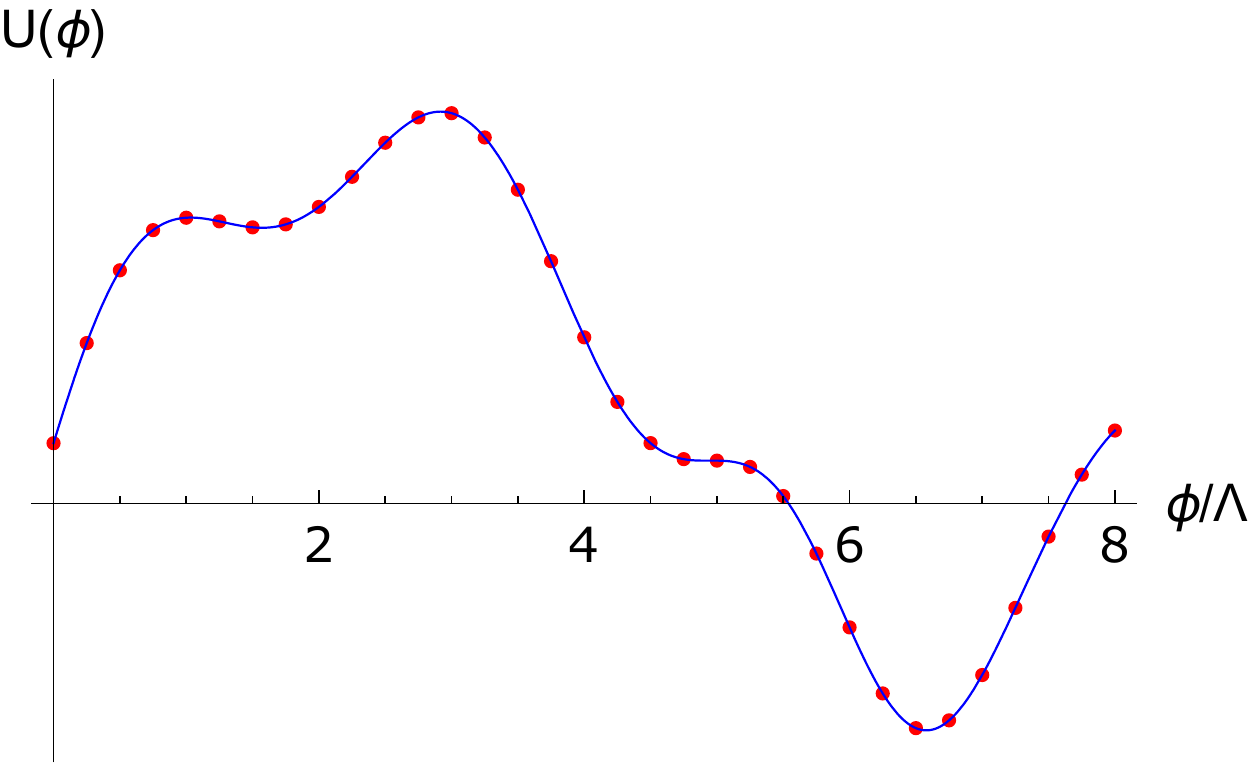}
  \includegraphics[width=2.2in]{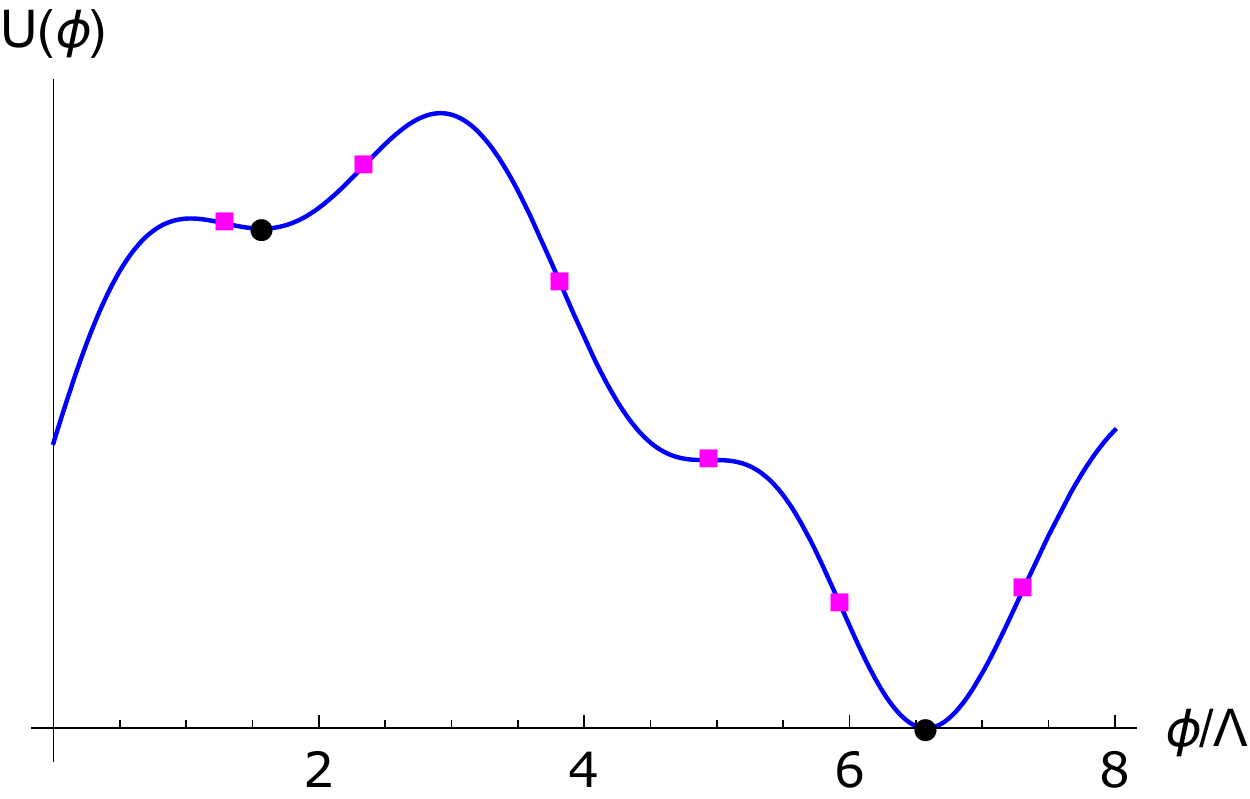}
  \includegraphics[width=2.2in]{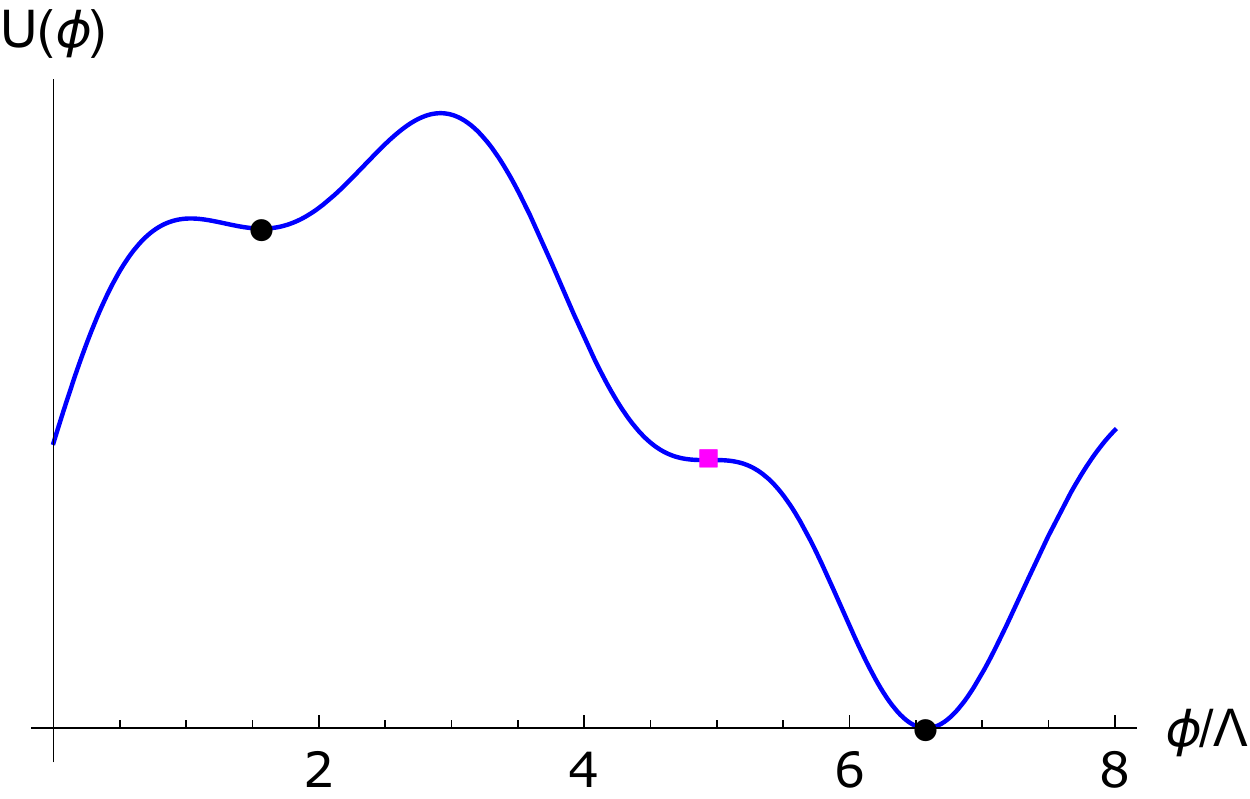}
\caption{ The procedure for choosing the relevant inflection and tunneling points as explained in the text.
Potential values generated by the random Gaussian algorithm (upper left frame) are interpolated by a smooth curve (upper right frame).
 In the lower left frame the minima and the inflection points are marked by blue dots and magenta squares, respectively, and the potential is shifted so that $U_{\rm TV} =0$.  The lower right frame shows the inflection point satisfying the selection criteria.}
\label{fig:procedure}
\end{figure}

In each realization of $U(\phi)$, we identify all extrema and inflection points.  For any pair of adjacent minima, we refer to the higher- and lower-energy ones as $\phi_{\rm FV}$ and $\phi_{\rm TV}$, respectively, and shift the potential so that $U_{\rm TV}=0$.  We keep only realizations that have an inflection point $\phi_{\rm I}$ satisfying the following criteria: (i) it is located between the top of the barrier and the true vacuum, (ii) its energy density is lower than that of the false vacuum, $U_{\rm I}<U_{FV}$. 
This procedure is illustrated in Fig.~\ref{fig:procedure}.

For each of the selected pairs of vacua, we seek instanton solutions of Eq.~(\ref{phibareq}) using the
shooting method.  We then use Eq.~(\ref{phi0}) to find the initial value(s) of the field $\phi$ after
tunneling.

\subsection{Distributions for $\phi_0$ and $N_{\rm max}$}

\begin{figure}[t] 
   \centering
   \includegraphics[width=2.5in]{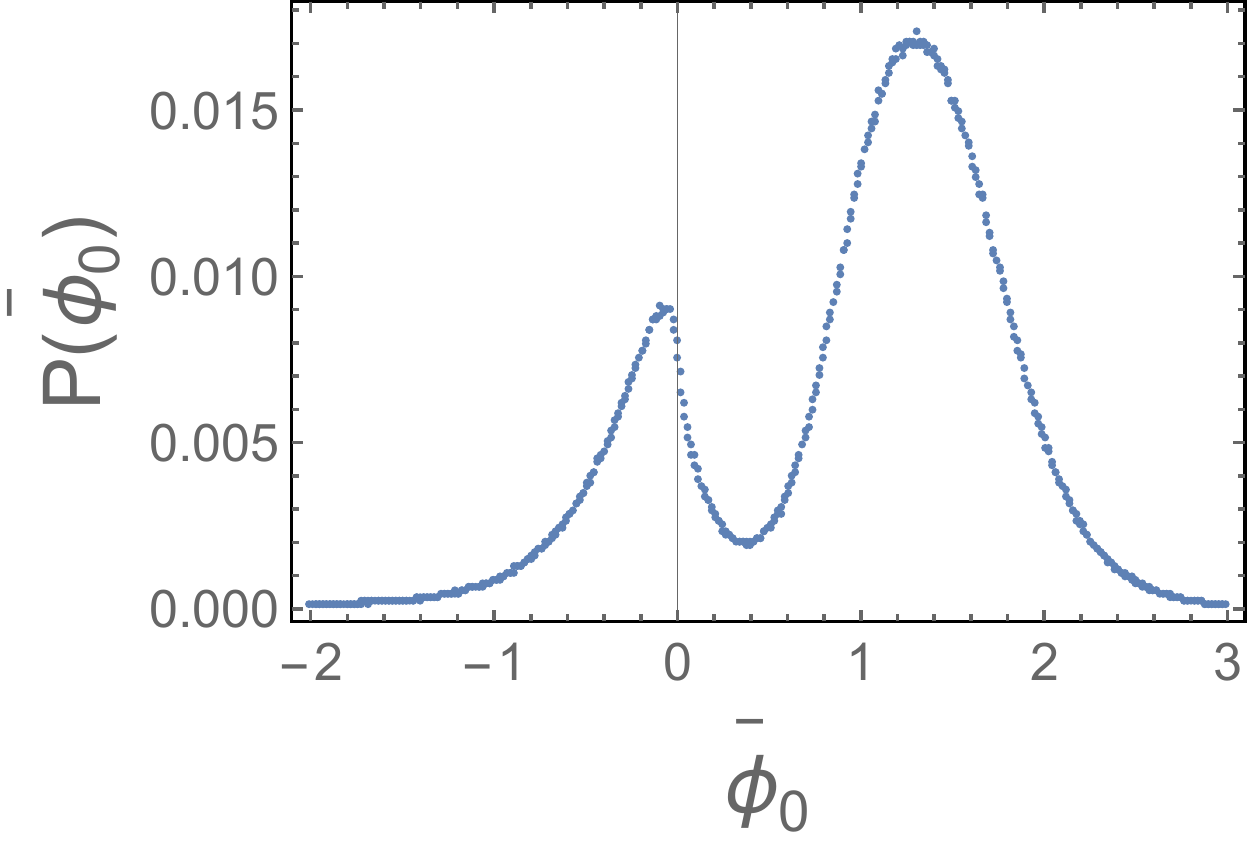} 
   \qquad 
   \includegraphics[width=2.5in]{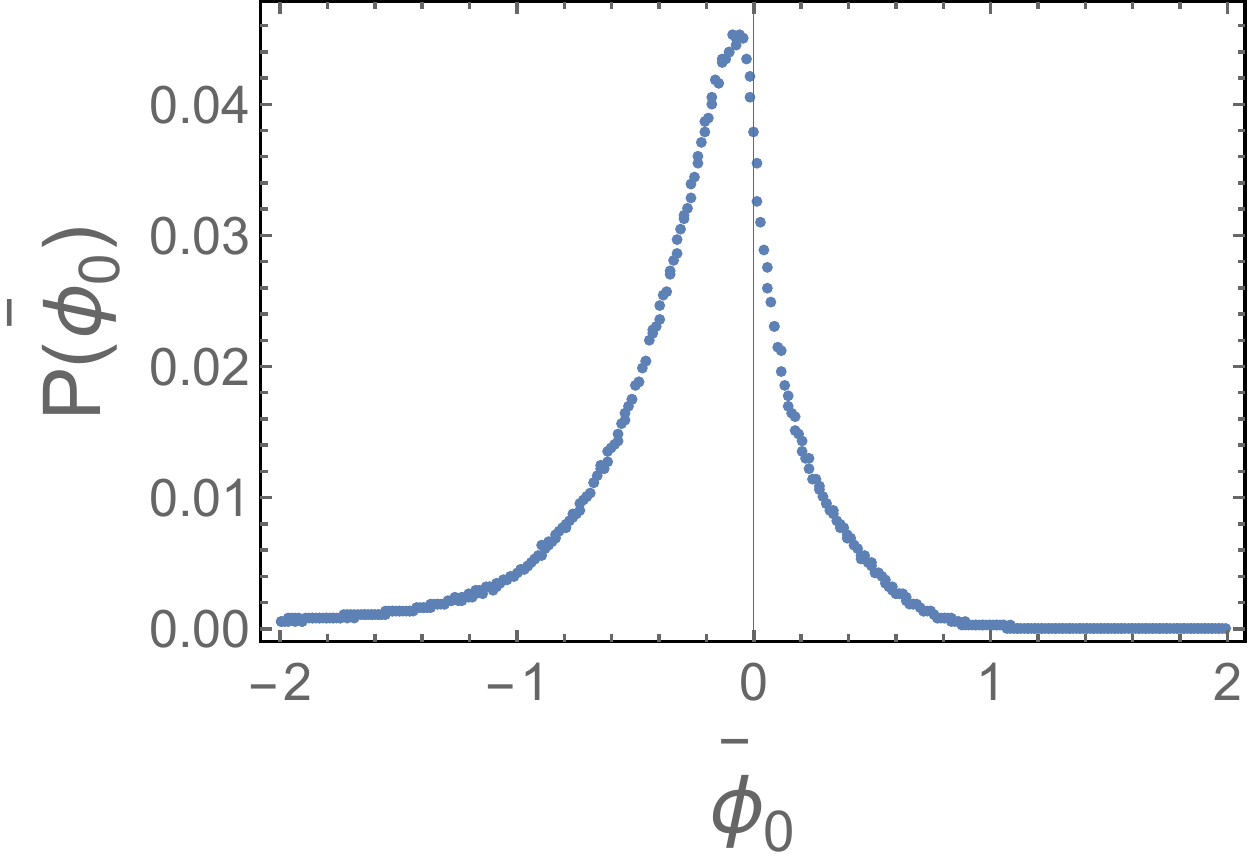}    
   \caption{Histogram of $\bar{\phi}_0$ for all tunneling processes (left panel) and for 
    the tunnelings to the inflection point (right panel). 
}
   \label{fig:histogram1}
\end{figure}

We plotted the distribution of the initial values ${\bar\phi}_0=\phi_0/\Lambda$ in the left panel of Fig.~\ref{fig:histogram1}.  Here and hereafter, the normalization of probability distributions is arbitrary.  In cases where the potential admitted two instanton solutions, we included the values of $\bar{\phi}_0$ for both of them (disregarding the subdominant ``middle'' instanton).  The distribution has a sharp peak centered near ${\bar\phi}_0=-0.1$ and a somewhat larger and broader peak at ${\bar\phi}_0\sim 1.3$.  These peaks correspond to tunnelings to the vicinity of the inflection point $\bar{\phi}_{\rm I} =0$ and of the true vacuum $\bar{\phi}_{TV}$, respectively.  In the right panel of Fig.~\ref{fig:histogram1} we plotted the distribution of ${\bar\phi}_0$ for tunnelings to the inflection point -- that is, including only cases where $\bar{\phi}_0$ is closer to $\bar{\phi}_{\rm I} = 0$ than to $\bar{\phi}_{TV}$.  This distribution is peaked near ${\bar\phi} \approx -0.1$ with a width $\Delta{\bar\phi} \approx 0.4$.

\begin{figure}[t] 
   \centering
   \includegraphics[width=2.5in]{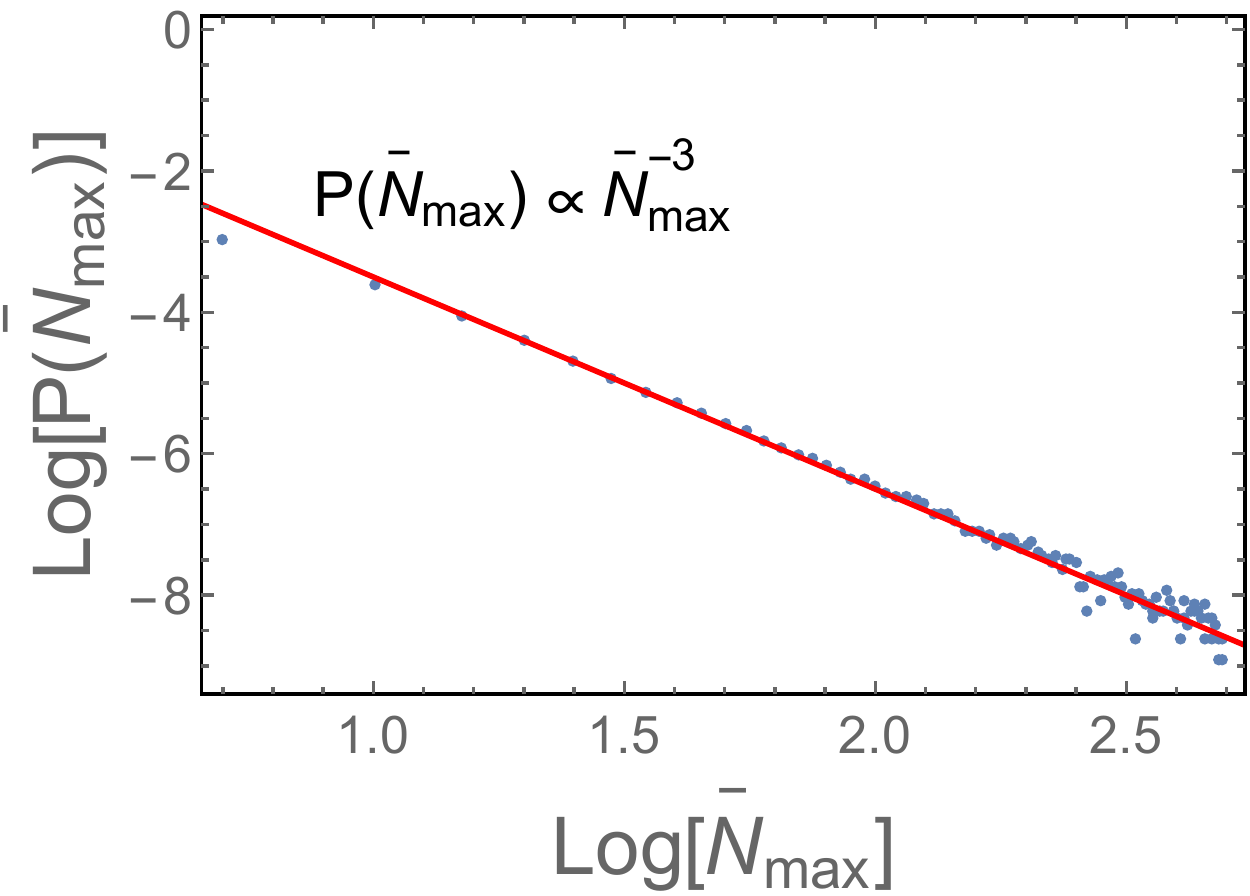}    
   \caption{
   Histogram of $\bar{N}_{\rm max}$ for all eligible inflection points. 
   The red line is an analytic function proportional to $\bar{N}_{\rm max}^{-3}$. 
}
   \label{fig:histogram2}
\end{figure}

Fig.~\ref{fig:histogram2} shows a histogram of the maximal e-fold number $\bar{N}_{\rm max}$, evaluated from Eq.~(\ref{Nmax}).  The result is well fitted by the analytic function
\beq 
P(\bar{N}_{\rm max}) \propto \bar{N}_{\rm max}^{-3},
\label{PNbar}
\eeq
which is shown by a red line in the figure.  This shows that the form of the $\bar{N}_{\rm max}$ distribution is not affected by the shift of the potential to $U_{\rm TV}=0$.

\begin{figure}[t] 
   \centering
   \includegraphics[width=2.5in]{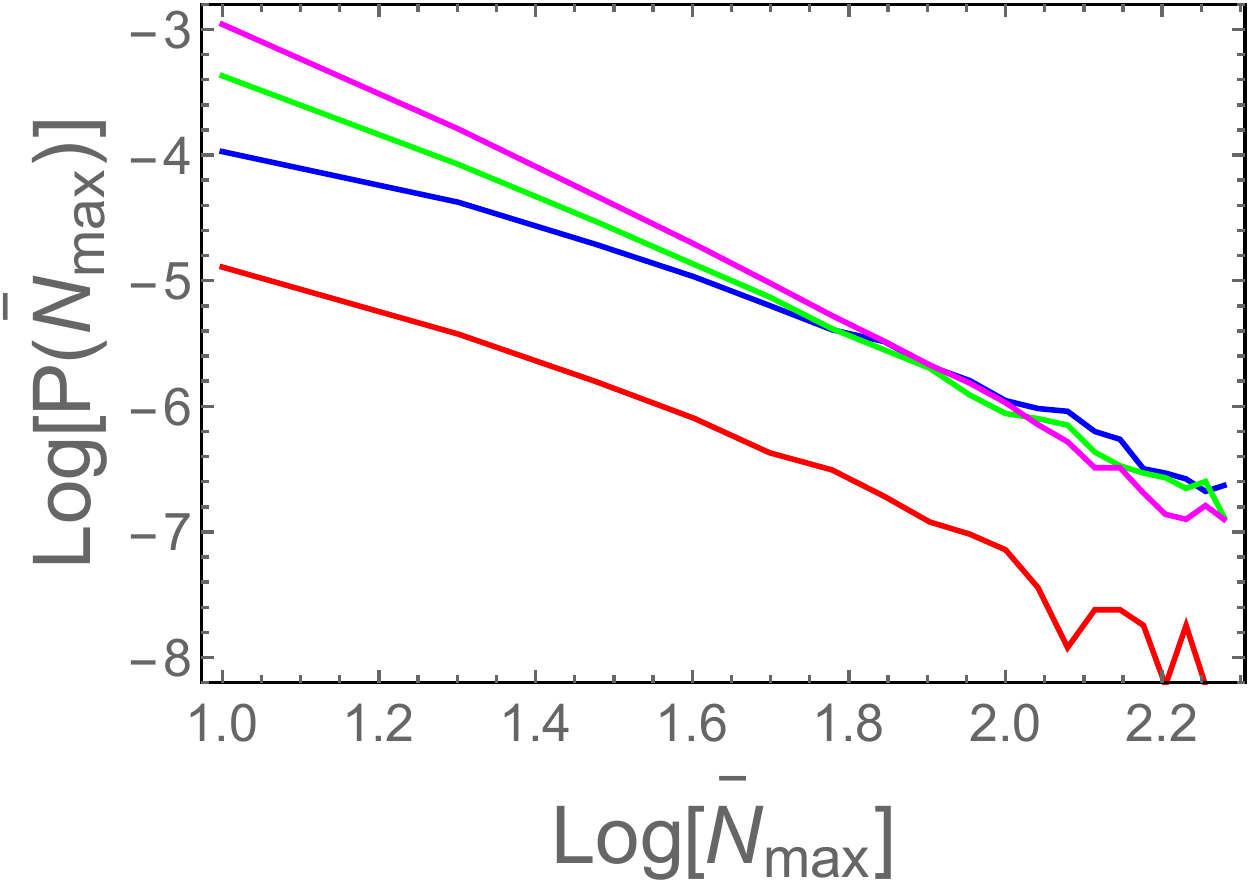} 
   \qquad 
      \includegraphics[width=2.5in]{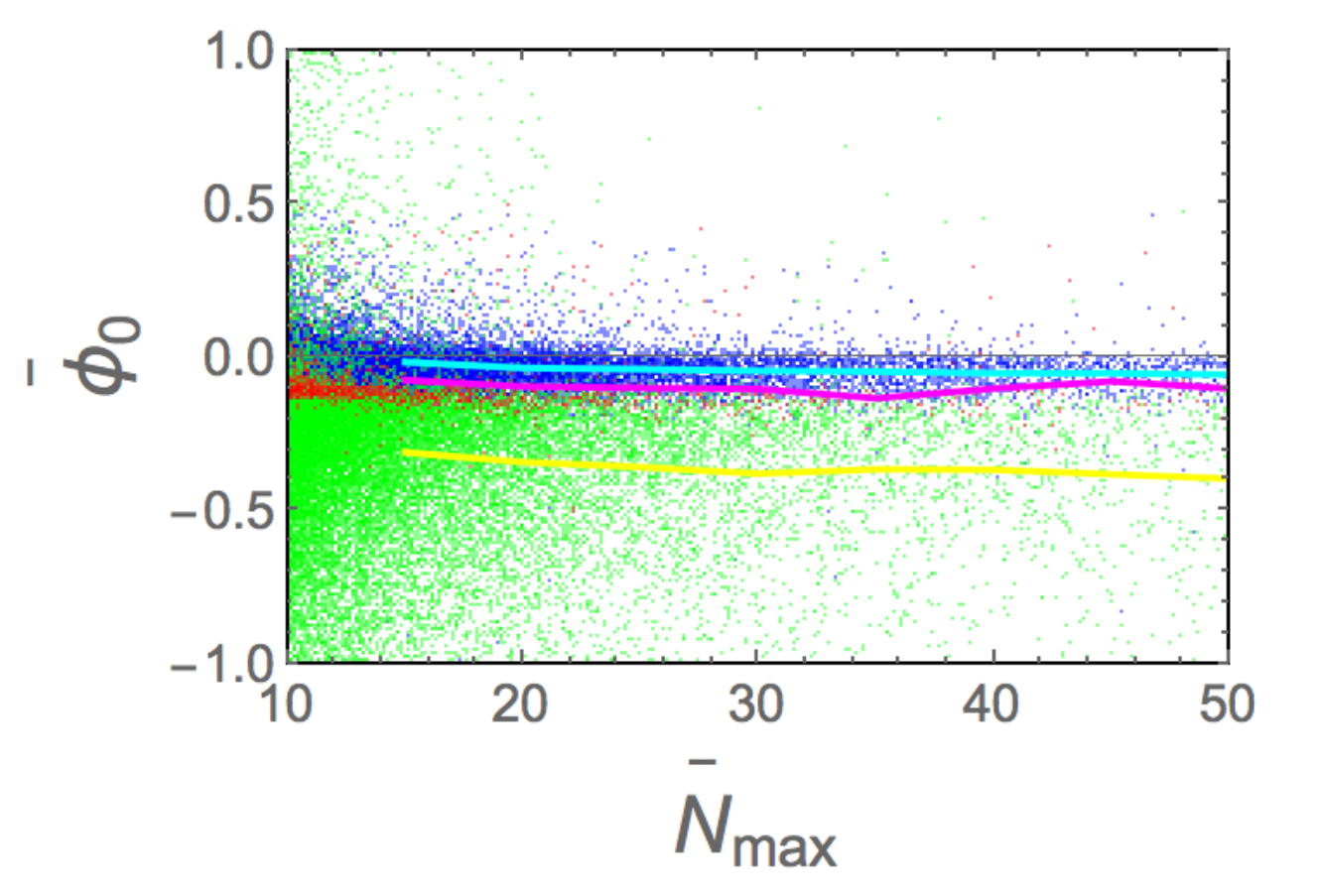} 
   \caption{The left panel shows the distribution of $\bar{N}_{\rm max}$ for the single instanton case with tunneling to the inflection point (green) and to the true vacuum (magenta) and for the multiple instanton case with $\bar{S}_{\rm T} > \bar{S}_{\rm I}$ (red) and $\bar{S}_{\rm T} < \bar{S}_{\rm I}$ (blue).  The right panel is a scatter plot in the ${\bar N}_{\rm max}-\phi_0$ plane with the same color code.  The yellow, magenta, and cyan lines represent the averaged values of $\bar{\phi}_0$.   
}
   \label{fig:histogram4}
\end{figure}

The plot in the left panel of Fig.~\ref{fig:histogram4} is the same as in Fig.~\ref{fig:histogram2}, but
now the contributions due to different types of instantons are shown separately.  The green (magenta)
lines represent the cases where there is only one instanton solution and $\bar{\phi}_0$ is closer to the
inflection point (true vacuum).  The red (blue) lines are for realizations with multiple instantons,
where the dominant tunneling is to the inflection point (true vacuum).  The type of instanton is not
relevant for the present paper, but our results may be useful in other contexts, so we present them here
for completeness. The figure shows that in the multiple instanton case the dominant tunneling channel is
almost always to the true vacuum: the number of realizations with $S_{\rm I}<S_{\rm T}$ is suppressed by
more than an order of magnitude.  This is consistent with the discussion of the tunneling action in
Sec.~\ref{sec:general formalism}.  We note also that the numbers of landscape realizations represented by
blue and green curves are nearly the same and are within about a factor of 2 from those represented by
the magenta curve.  We have not found any explanation for this surprising fact.  It implies that
realizations with large values of $N_{\rm max}$ split into three comparable groups: a group with
tunneling only to inflection point, a group with tunneling to the true vacuum, and a group with multiple
tunneling channels, where the tunneling to the true vacuum dominates.

The right panel of Fig.~\ref{fig:histogram4} is a scatter plot of realizations in
$\bar{\phi}_0$-$\bar{N}_{\rm max}$ plane, with the same color code as in the left panel.  The average
values $\langle\bar{\phi}_0\rangle$ are shown as yellow, magenta, and cyan lines for the data indicated
by green, red, and blue dots, respectively.  We see that $\langle\bar{\phi}_0\rangle$ is almost constant
at $\bar{N}_{\rm max} \gg 1$ for all types of instantons, with $\langle\bar{\phi}_0\rangle \sim - 0.4$
for the single instanton case and $\langle\bar{\phi}_0\rangle\sim - 0.1$ for the multi-instanton case.
This indicates that ${\bar N}_{\rm max}$ and ${\bar\phi}_0$ are essentially uncorrelated in the most
interesting regime of large ${\bar N}_{\rm max}$.

From now on, we shall not distinguish between single and multi-instanton tunnelings and treat all
inflection-point tunnelings on equal footing.  In multi-instanton realizations with a dominant
true-vacuum instanton, we keep only the inflection-point instanton.  The reason is that true-vacuum
tunneling is irrelevant for our discussion, and the tunneling processes described by inflection-point
instantons occur regardless of whether or not there is a more probable tunneling channel.

\subsection{Tunneling action}
\label{sec:tunneling action}

The distributions for $\phi_0$ and $N_{\rm max}$ that we calculated here are defined as frequencies of
occurrence in the landscape.  They should not be confused with the probabilities of occurrence in the
multiverse, which can be directly related to observational predictions.  The problem of defining these
probabilities is known as the measure problem, which at present remains unresolved. (For a review of the
measure problem see \cite{Freivogel}.)  A number of different measure prescriptions have been suggested
in the literature.  Some of them lead to paradoxes or to a glaring conflict with observations and have
therefore been ruled out.  This process of elimination has not been enough to fix a unique measure of the
multiverse.  However, the measure prescriptions which are not obviously problematic tend to give similar
predictions and introduce similar weighting factors for different realizations of the potential.  The
scale factor measure \cite{scalefactor1,scalefactor2} can be taken as a representative example of such
``acceptable'' measures.

For a given measure prescription, probabilities can be calculated by solving the rate equation, which is
similar to the Boltzmann equation in the multiverse.  Naively, one might expect that different tunneling
realizations in the landscape should be weighted by the tunneling rate, which is proportional to
$\exp(-S_E)$, where $S_E$ is the tunneling action.  However, analysis of the rate equation in the
scale-factor measure shows that this expectation is incorrect.  A simple counter-example is a landscape
with an everywhere positive potential, $U(\phi)>0$.  It can be shown that in such a landscape the
probabilities depend only on the vacuum energy density and are independent of the transition rates
\cite{Vanchurin}.  In general, the probability of a given vacuum has a complicated dependence on the
transition rates between different vacua in the landscape, not just on the rate of tunneling to this
particular vacuum.  One also finds that transitions with a small tunneling action are not generally
``rewarded'' with a high weighting factor.  The reason can be roughly explained as follows.  The
weighting factor for vacuum $i$ due to tunneling from a false vacuum $j$ is proportional to $\kappa_{ij}
f_j$, where $\kappa_{ij}$ is the tunneling rate from $j$ to $i$ per Hubble volume per Hubble time, and
$f_j$ is the volume fraction occupied by vacuum $j$ on constant scale factor surfaces.  If the tunneling
rate is very high, this leads to a rapid depletion of the false vacuum, so $f_j$ gets very small.  These
two effects tend to compensate one another.  On the other hand, tunnelings with a large instanton action
tend to be disfavored in the presence of other decay channels with a smaller action.

A study of the rate equation in a random Gaussian landscape would require a complicated statistical
analysis, which is beyond the scope of the present paper.  To facilitate such analysis in future work,
here we shall analyze possible correlations of the action ${\bar S}$ with ${\bar\phi}_0$ and ${\bar
  N}_{\rm max}$.  If present, such correlations may affect probabilities for the initial conditions and
for the observational effects of inflation.

\begin{figure}[t] 
   \centering
   \includegraphics[width=2.5in]{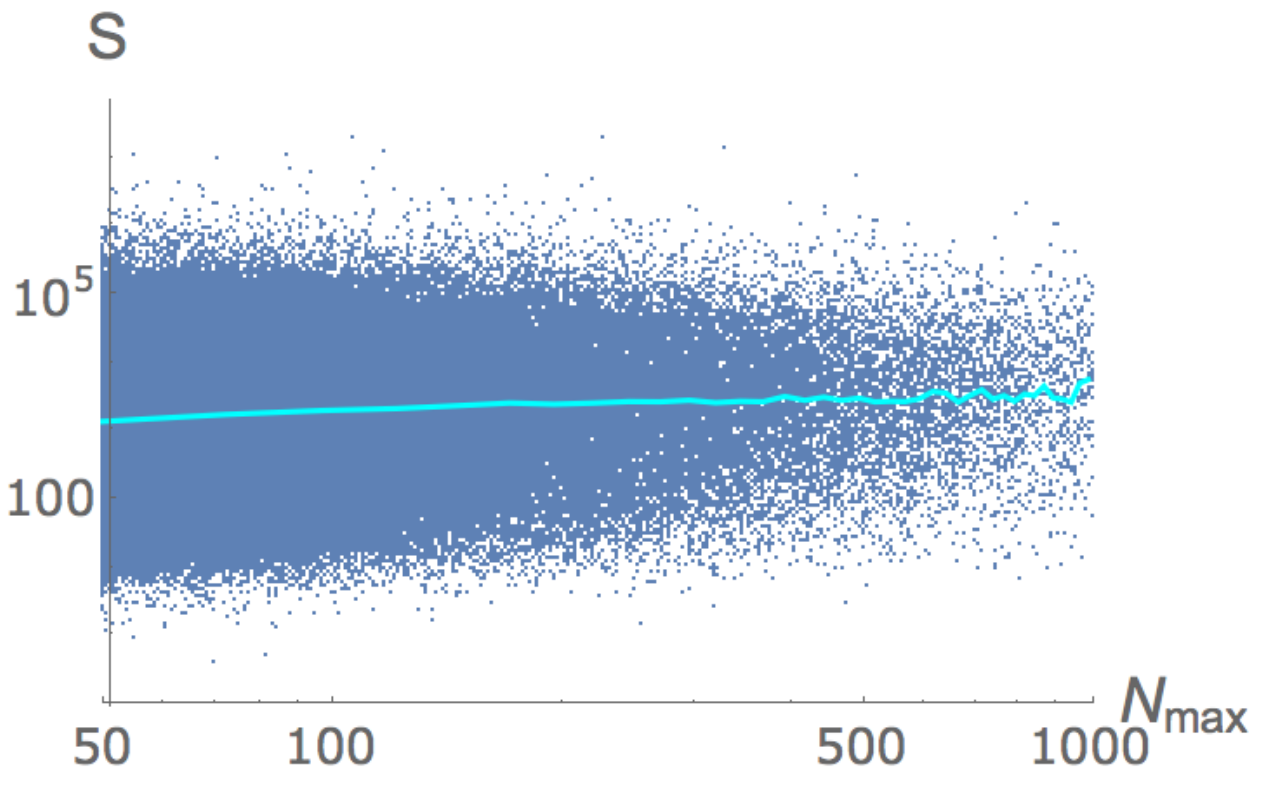} 
   \qquad 
   \includegraphics[width=2.5in]{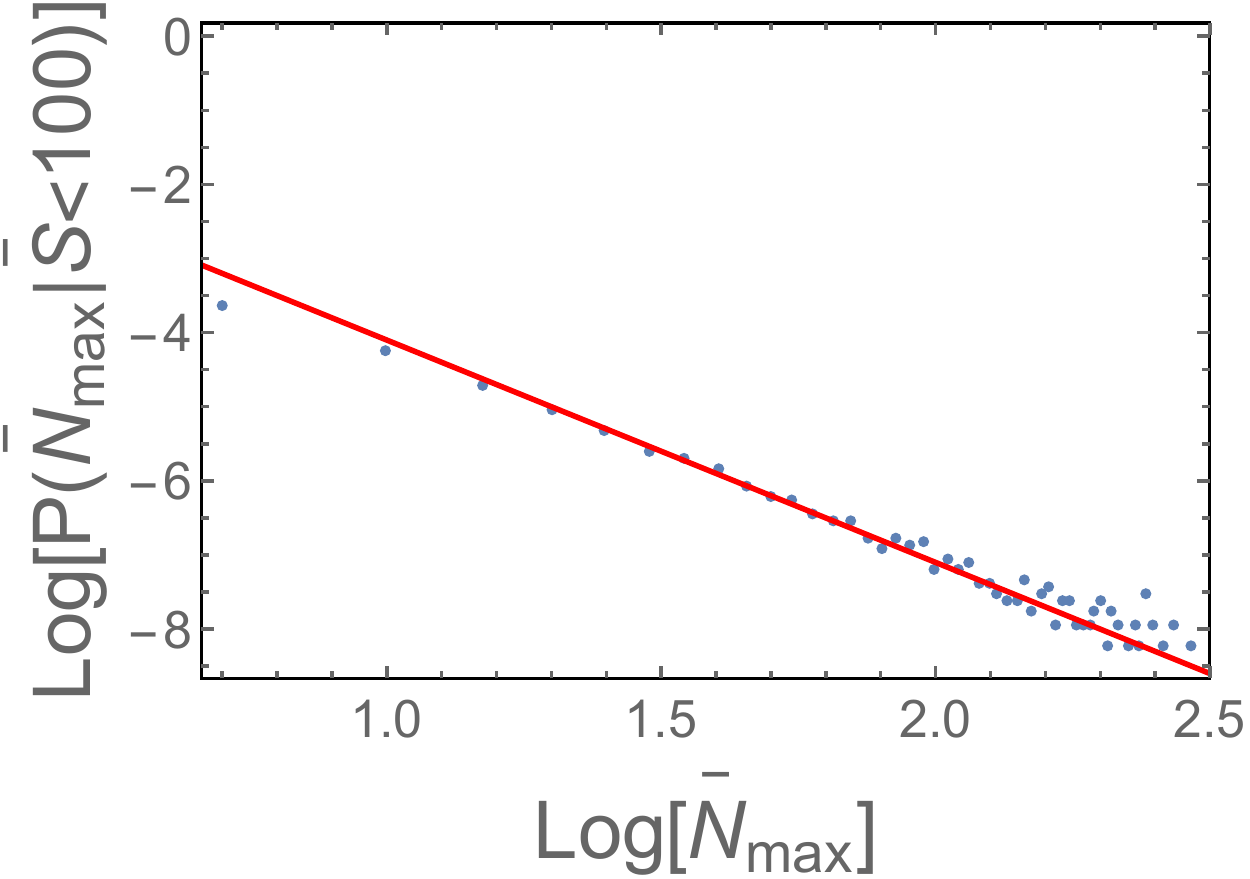}    
   \caption{
   Scatter plot of ${\rm Log} [\bar{S}]$ and ${\rm Log} [\bar{N}_{\rm max}]$ (left panel) 
   and the probability distribution of ${\rm Log} [\bar{N}_{\rm max}]$ under the condition of $\bar{S} < 100$ (right panel).  The left panel also shows the average value of $\bar{S}$ as a function of ${\bar N}_{\rm max}$, and the right panel shows the function ${\bar N}_{\rm max}^{-3}$.
}
   \label{fig:action2}
\end{figure}

We show a scatter plot of ${\rm Log} [\bar{S}]$ and ${\rm Log} [\bar{N}_{\rm max}]$ in the left panel of
Fig.~\ref{fig:action2}.  This includes only tunnelings to the inflection point.  For large values of
$\bar{N}_{\rm max}$, the rescaled instanton action is mostly in the range $100 \lesssim{\bar S}\lesssim
10^4$.  Note that the full action (\ref{rescaleS}) is much larger than that.  From Eq.~(\ref{ULambda}) we
have \beq \frac{\Lambda^4}{U_0} \gg 10^{12}, \eeq and thus $S_E \gg 10^{14}$.  The plot suggests that for
$\bar{N}_{\rm max}\gg 1$, $\bar{S}$ is essentially uncorrelated with $\bar{N}_{\rm max}$.  In particular,
the average value of ${\rm Log} [\bar{S}]$ is nearly constant at large $\bar{N}_{\rm max}$.  This is not
surprising, since the instanton action depends on the shape of the potential around the top of the
barrier, while ${\bar N}_{\rm max}$ is not sensitive to this shape.  The right panel of
Fig.~\ref{fig:action2} shows the probability distribution of $\bar{N}_{\rm max}$ under the condition of
$\bar{S}< 100$.  The distribution is still well fitted by $N_{\rm max}^{-3}$.

\begin{figure}[t] 
   \centering
   \includegraphics[width=2.5in]{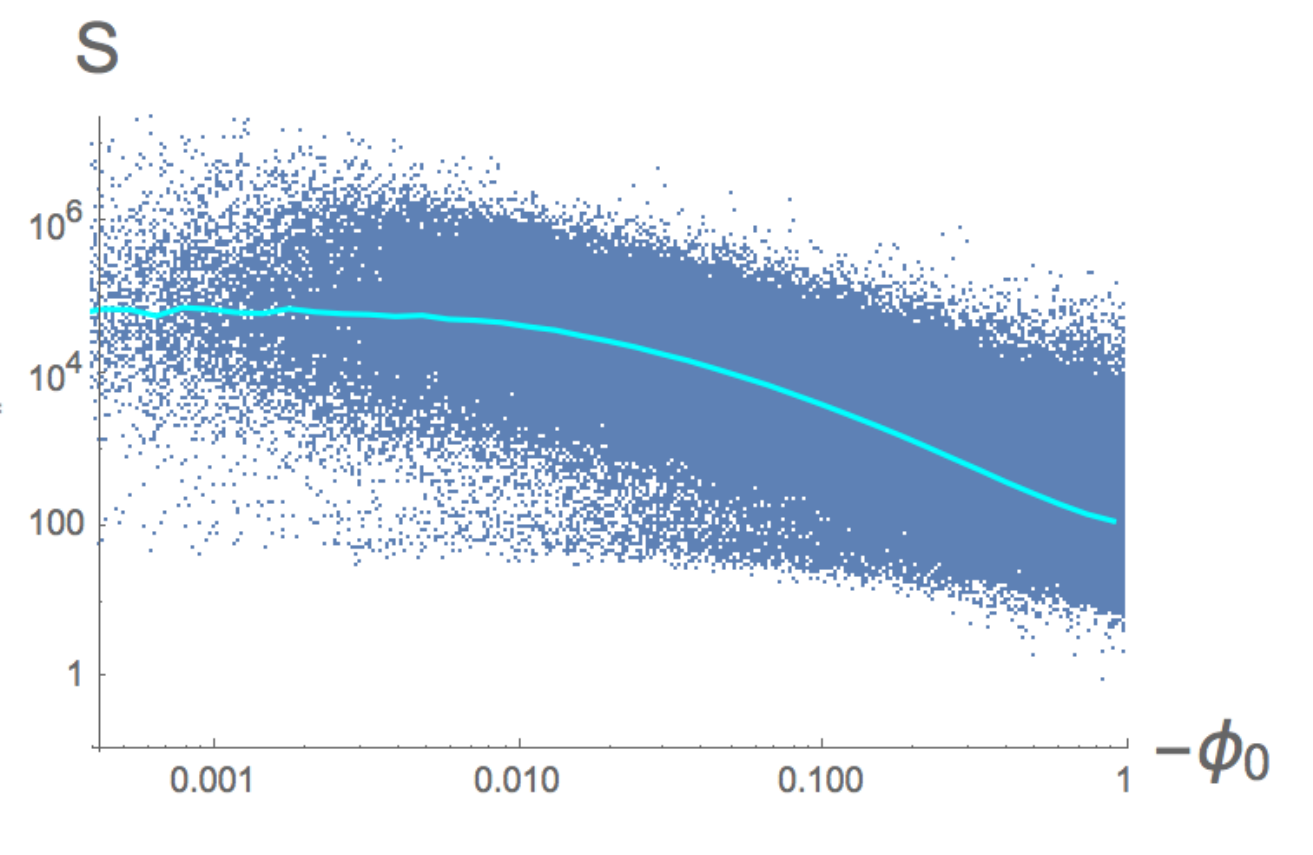} 
   \caption{
   Scatter plot of ${\rm Log} [\bar{S}]$ and ${\rm Log} [-\bar{\phi}_0]$  
   under the condition of $\bar{N}_{\rm max} >30$. 
   }
   \label{fig:action3}
\end{figure}

Figure~\ref{fig:action3} is a scatter plot in the ${\rm Log} [\bar{S}]$-${\rm Log} [-\bar{\phi}_0]$ plane
for $\bar{N}_{\rm max} >30$.  It exhibits significant correlation between $\bar{S}$ and $\bar{\phi}_0$.
In particular, the number of realizations with large values of ${\bar S}$ increases towards small values
of $\bar{\phi}_0$. This can be understood as the contribution of realizations with $U_{\rm FV} \approx
U_{\rm I}$, which correspond to the thin-wall regime.  In the limit $U_{\rm FV} \to U_{\rm I}$, the
tunneling action diverges and $\bar{\phi}_0 \to 0$ \cite{Coleman}.  On the other hand, the average value
of S appears to saturate at a constant $\langle {\bar S} \rangle \sim 10^5$ at ${\bar\phi}\to 0$.  In the
rest of the paper we shall assume that the correlations between ${\bar S}$ and ${\bar\phi}_0$ have no
significant effect on the probability distribution for $\phi_0$. This issue, however, requires some
further study.

\section{Slow-roll inflation after tunneling} \label{sec:1D}

After tunneling, the bubble has the interior geometry of an open FRW universe,  
\beq
ds^2 = dt^2-a^2(t)\left(d\chi^2+\sinh^2\chi d\Omega^2\right).
\eeq
Its evolution is described by the equations
\beq
\frac{{\dot a}^2}{a^2}=\frac{1}{3}\left(U(\phi)+\frac{{\dot\phi}^2}{2}\right)+\frac{1}{a^2},
\label{Friedmann}
\eeq
\beq
{\ddot\phi}+3\frac{\dot a}{a}{\dot\phi}+U'(\phi)=0,
\label{phiU}
\eeq
with the initial conditions at $t=0$
\beq
{\dot a}(0)=1,~~~ \phi(0)=\phi_0,~~~ {\dot\phi}(0)=0,
\eeq
where $\phi_0$ is determined from the instanton solution.
Our aim in this section is to find the probability distribution for the e-folding number after tunneling $N_e$.

We first note that the results of Sec.~\ref{sec:initial condition} for the initial value ${\bar\phi}_0 = \phi_0/\Lambda$ are independent of $\Lambda$ and can be applied for any value of $\Lambda\ll 1$.  Thus we expect the tunneling to occur to a point $\phi_0$ with a probability distribution of width 
\beq
\Delta\phi_0 \sim 0.4 \Lambda, 
\label{tunnelingrange}
\eeq
near the inflection point (which we assume to be at $\phi=0$).  We note also that the range of $\phi$ around the inflection point where slow-roll inflation is possible, $|\phi|\lesssim \Lambda^3$, is much smaller than (\ref{tunnelingrange}) for small $\Lambda$.  Thus, most tunnelings will occur outside of the slow-roll range.

Furthermore, the dynamics of inflation does depend on the value of $\Lambda$.  In order to study this dynamics numerically, one would have to find a large sample of realizations of $U(\phi)$ with $N_{\rm max}\gg 1$.  But since $N_{\rm max}=\Lambda^2 \bar{N}_{\rm max}$, it follows from Eq.~(\ref{PNbar}) that the number of such realizations is suppressed by a factor $\Lambda^6$, so a sufficiently large sample can be found only for $\Lambda\gtrsim 0.3$.  We have therefore developed a semi-analytic approach to the problem.

The potential near the inflection point is given by Eq.~(\ref{cubic}),
\beq
 U(\phi) = U_{\rm I} + {U_{\rm I}}' \phi + \frac{1}{3!} {U_{\rm I}}''' \phi^3. 
 \label{cubic1}
\eeq
In the tunneling range (\ref{tunnelingrange}), the cubic term gives a correction to $U_{\rm I}$ of the order
\beq
\Delta U  \lesssim   (1 / 3!) {U_{\rm I}}''' (0.4 \Lambda)^3  \sim  10^{-2} U_0,   
\label{DeltaU}
\eeq
where in the last step we assume  ${U_{\rm I}}''' \sim U_0 /\Lambda^3$.  For $N_{\rm max}\gg 1$, the linear term in (\ref{cubic1}) is much smaller than the cubic term everywhere except in a small vicinity of $\phi = 0$.  Hence the potential after tunneling and until the end of inflation is well approximated by $U(\phi) \approx U_{\rm I}$.

The kinetic energy $(1/2){\dot\phi}^2$ does not exceed the cubic term (friction can only reduce it), so it is also negligible during inflation (compared to $U_{\rm I}$).  Thus the Friedmann equation (\ref{Friedmann}) can be approximated as
 \beq
{\dot a}^2 = 1 + H_{\rm I}^2 a^2,
\eeq
where $H_{\rm I}^2 = U_{\rm I} /3$.  The solution is
\beq
a(t) = H_{\rm I}^{-1} \sinh (H_{\rm I} t).  
\label{at}
\eeq
This implies that inflation starts at $t \sim H_{\rm I}^{-1}$, after a brief curvature-dominated period. 

With the scale factor (\ref{at}), Eq.~(\ref{phiU}) for $\phi(t)$ takes the form
\beq
{\ddot\phi} + 3{H_{\rm I}} \coth ({H_{\rm I}} t) {\dot\phi} + {U_{\rm I}}''' \phi^2 /2 + {U_{\rm I}}'  =  0.
\label{phieq1}
\eeq
Rescaling the variables as
\beq
{H_{\rm I}} t = \tau,  ~~~   (|{U_{\rm I}|}''' /(2 {H_{\rm I}}^2)) \phi = \psi,
\eeq
we have
\beq
{\ddot\psi} + 3\coth\tau {\dot\psi} - \psi^2  -q=  0,    
\label{psieq1}
\eeq
where dots now stand for derivatives with respect to $\tau$ and
\beq
q &=& \frac{{U_{\rm I}}' {U_{\rm I}}'''}{2 {H_{\rm I}}^4}, 
\\
&=& \frac{9 \pi^2}{N_{\rm max}^2} 
 \simeq
 0.01 \lmk \frac{N_{\rm max}}{100} \rmk^{-2}. 
\eeq
Here, $N_{\rm max} = \pi\sqrt{2} U_{\rm I}/\sqrt{{U_{\rm I}}' {U_{\rm I}}'''}$ is the maximal number of efolds defined by \eq{Nmax}.
The slow roll condition fails at the point where $\abs{U'' / U} = 1$, which means that the slow roll range is $\abs{\psi} < 3/2$. 
The initial conditions for $\psi(\tau)$ are
\beq 
\psi(0) \equiv\psi_0 = (|{U_{\rm I}}'''| /(2{H_{\rm I}}^2) ) \phi_0  \sim  \Lambda^{-3}\phi_0, ~~~ {\dot\psi}(0)=0.
\label{psi0}
\eeq

\subsection{Beginning of slow roll}

As we already noted, the tunneling range (\ref{psi0}) is much wider than the slow roll range $|\psi| \lesssim 1$ for small values of $\Lambda$.  If $\psi_0$ happens to be in this narrow range, then inflation begins at $t\sim H_{\rm I}^{-1}$, right after tunneling.  If $\psi_0 \gg 1$, then clearly inflation does not happen.  For $\psi_0 \ll -1$, the field starts rolling fast at $t\sim H_{\rm I}^{-1}$ and may overshoot part or all of the slow-roll region.  We shall find when (and whether) the slow roll begins assuming that the last term in Eq.~(\ref{psieq1}) can be neglected up to that moment.  This approximation is justified for large values of $N_{\rm max}$, as we shall later verify.

Without the last term, Eq.~(\ref{psieq1}) has no free parameters:
\beq
{\ddot\psi} + 3\coth\tau {\dot\psi} - \psi^2 =  0.    
\label{psieq2}
\eeq
Hence the only free parameter of the problem is the initial value $\psi_0$.   We solved Eq.~(\ref{psieq2})  
numerically to determine the value of $\psi$ at the onset of slow roll.

\begin{figure}[t] 
   \centering
  \includegraphics[width=2.5in]{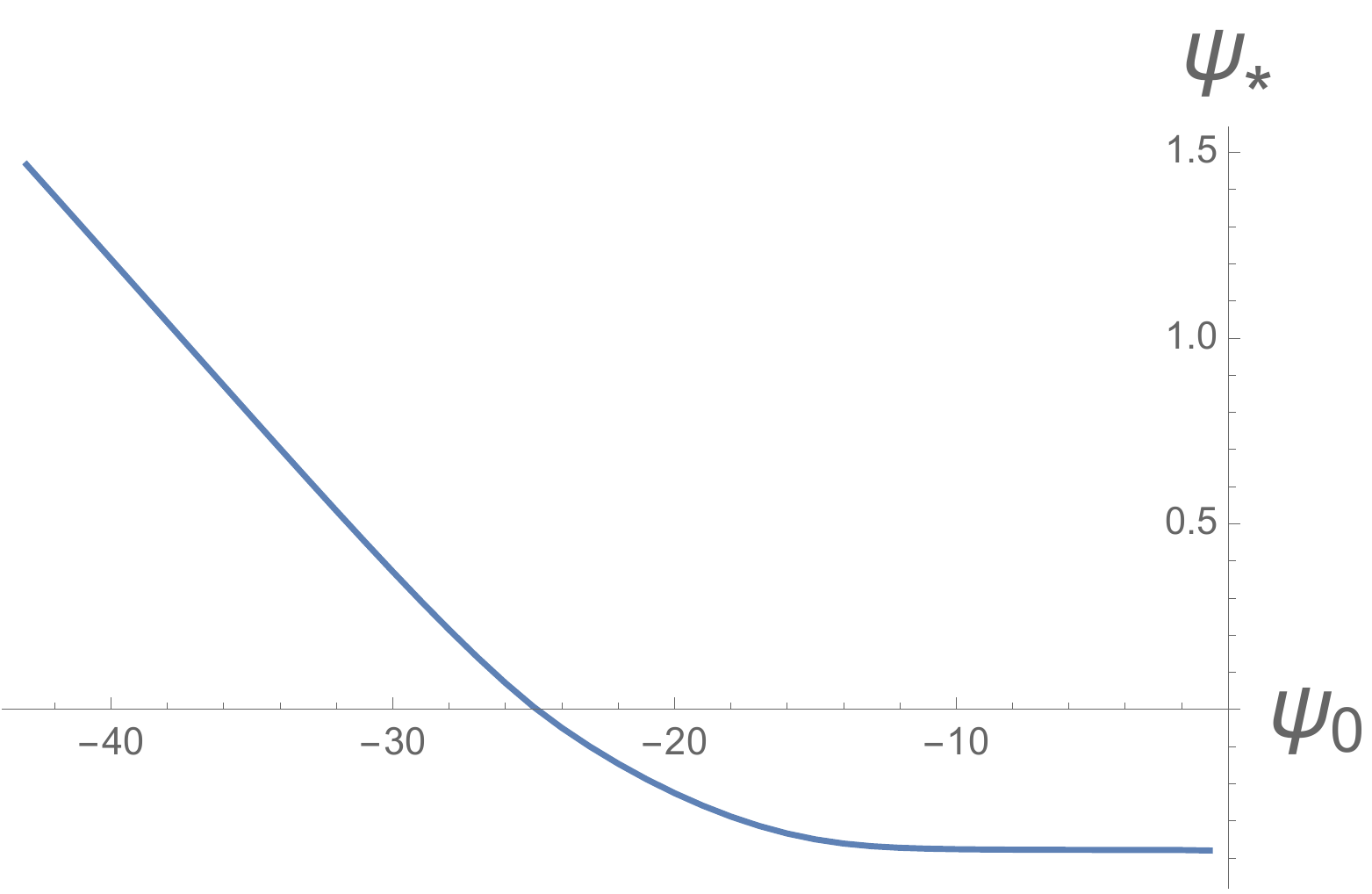}
  \caption{The value of $\psi_*$ as a function of $\psi_0$. 
}
   \label{fig:phi-star}
\end{figure}

If $\psi$ eventually gets into the slow-roll regime, the first term in Eq.~(\ref{psieq2}) becomes negligible compared to the other two terms; then
\beq
3\coth\tau {\dot\psi} \approx \psi^2 .
\eeq
For the purpose of our numerical analysis, we rewrite this condition as follows: 
\beq
 \left\vert \frac{{\dot\psi} - \psi^2/(3\coth \tau) }{{\dot\psi}} \right\vert < 0.1. 
\label{cond2}
\eeq
The value of $\psi$ when this is first satisfied marks the beginning of slow roll.  We shall denote it by $\psi_*$.  In Fig.~\ref{fig:phi-star} we plotted $\psi_*$ as a function of $\psi_0$.

For $\psi_0 < -42$, the condition (\ref{cond2}) is never satisfied, indicating that the inflaton field overshoots the entire slow-roll region.  We note also that there is a critical value of $\psi_0$, $\psi_{\rm cr}\simeq -24.8$, above which $\psi_*$ is negative and slow-roll starts before the inflaton reaches the inflection point.   For later use we also give the slope of the curve in Fig.~\ref{fig:phi-star} at $\psi_0 =\psi_{\rm cr}$: 
\beq
\left. \frac{\dd \psi_*}{\dd \psi_0} \right\vert_{\psi_0 \simeq \psi_{\rm cr}} \simeq -0.06 . 
\label{psi'}
\eeq

\subsection{The number of e-folds}

The number of e-folds of slow-roll inflation can now be found from
\beq
N_e \approx -\int_{\phi_*}^{\phi_e} d\phi\frac{U(\phi)}{U'(\phi)},
\eeq
where $\phi_*=(2U_{\rm I}/3|{U_{\rm I}'''}|) \psi_*$ is the value of $\phi$ corresponding to $\psi_*$.  As before, we replace the upper bound of integration by $\infty$ and approximate $U(\phi)$ by $U_{\rm I}$ in the numerator.  However, we can no longer neglect the linear term in the potential, since otherwise the integral would diverge at $\phi=0$.  Thus we obtain
\beq
N_e \approx -U_{\rm I} \int_{\phi_*}^{\infty} \frac{d\phi}{{U_{\rm I}'}+\frac{1}{2}{U_{\rm I}}''' \phi^2}
=\frac{N_{\rm max}}{2}\left[1-\frac{2}{\pi} \arctan\left(\frac{N_{\rm max}}{3\pi} \psi_* \right)\right]. 
\label{Ne-psistar}
\eeq

For $\psi_* <0$ and $|\psi_*| \gg 3\pi/N_{\rm max}$, we can use $\arctan (-x) \approx -\pi/2 + x^{-1}$, which gives
\beq
N_e\approx N_{\rm max}-\frac{3}{|\psi_*|}.
\label{Ne2}
\eeq
With $N_{\rm max}\gtrsim 100$, the second term in (\ref{Ne2}) can be significant only for $|\psi_*| \lesssim 0.1$.  From the graph in Fig. \ref{fig:phi-star} we see that this is satisfied only in a narrow range $\Delta\psi_0 \sim 1$ around $\psi_0 =\psi_{\rm cr}$.  We thus conclude that $N_e \approx N_{\rm max}$ in most of the range 
\beq
\psi_{\rm cr} < \psi_0 \lesssim -1.
\label{target}
\eeq
We shall call it the target range.  

Similarly, for positive $\psi_* \gg 3\pi/N_{\rm max}$ we find
\beq
N_e\approx \frac{3}{\psi_*}.
\label{Ne3}
\eeq
In order to have $N_e \gtrsim 50$, we need $\psi_*\lesssim 0.1$, and once again this is satisfied only in a narrow range near $\psi_{\rm cr}$.

\begin{figure}[t] 
   \centering
   \includegraphics[width=2.5in]{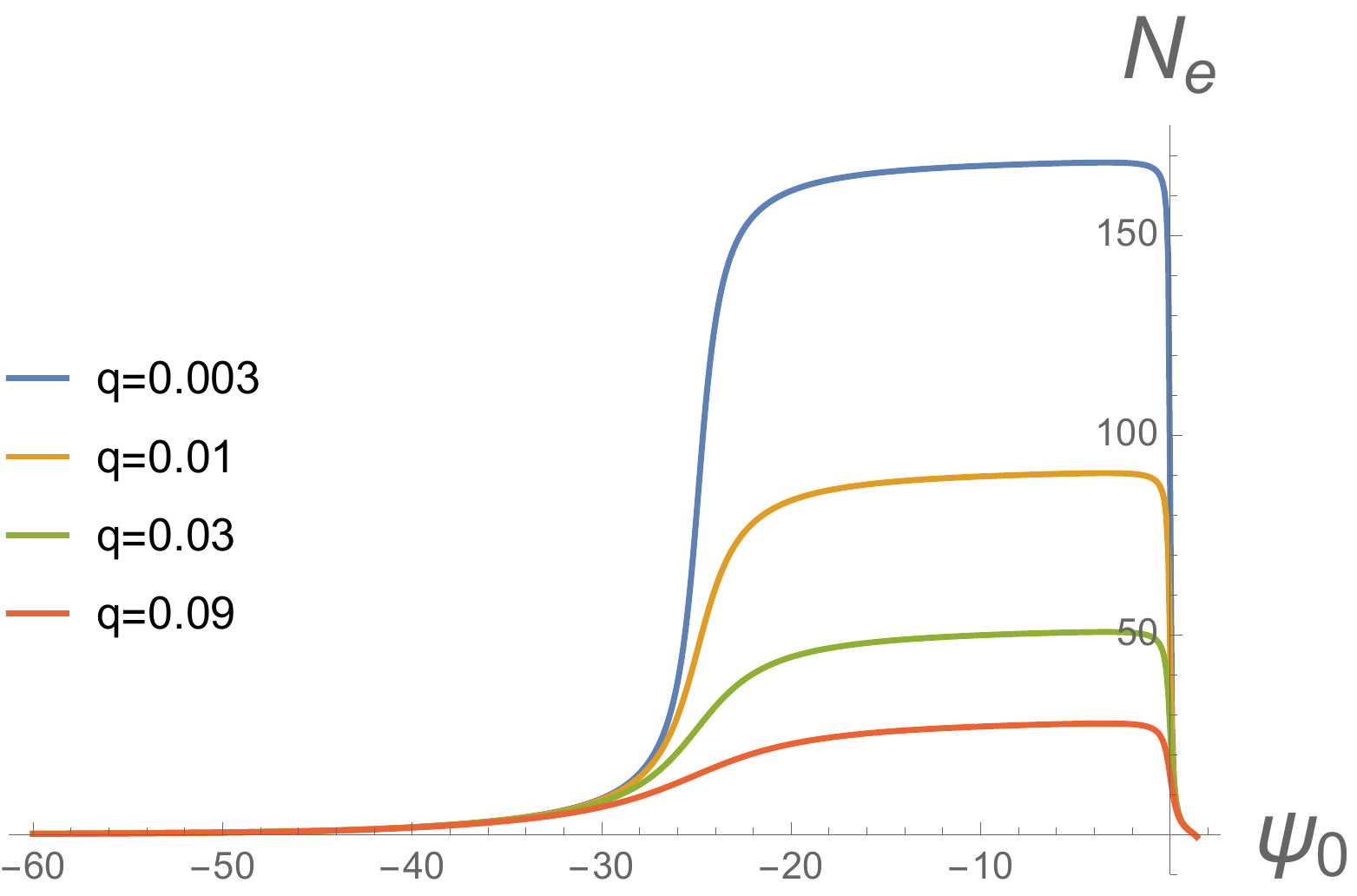}    
      \qquad
   \includegraphics[width=2.5in]{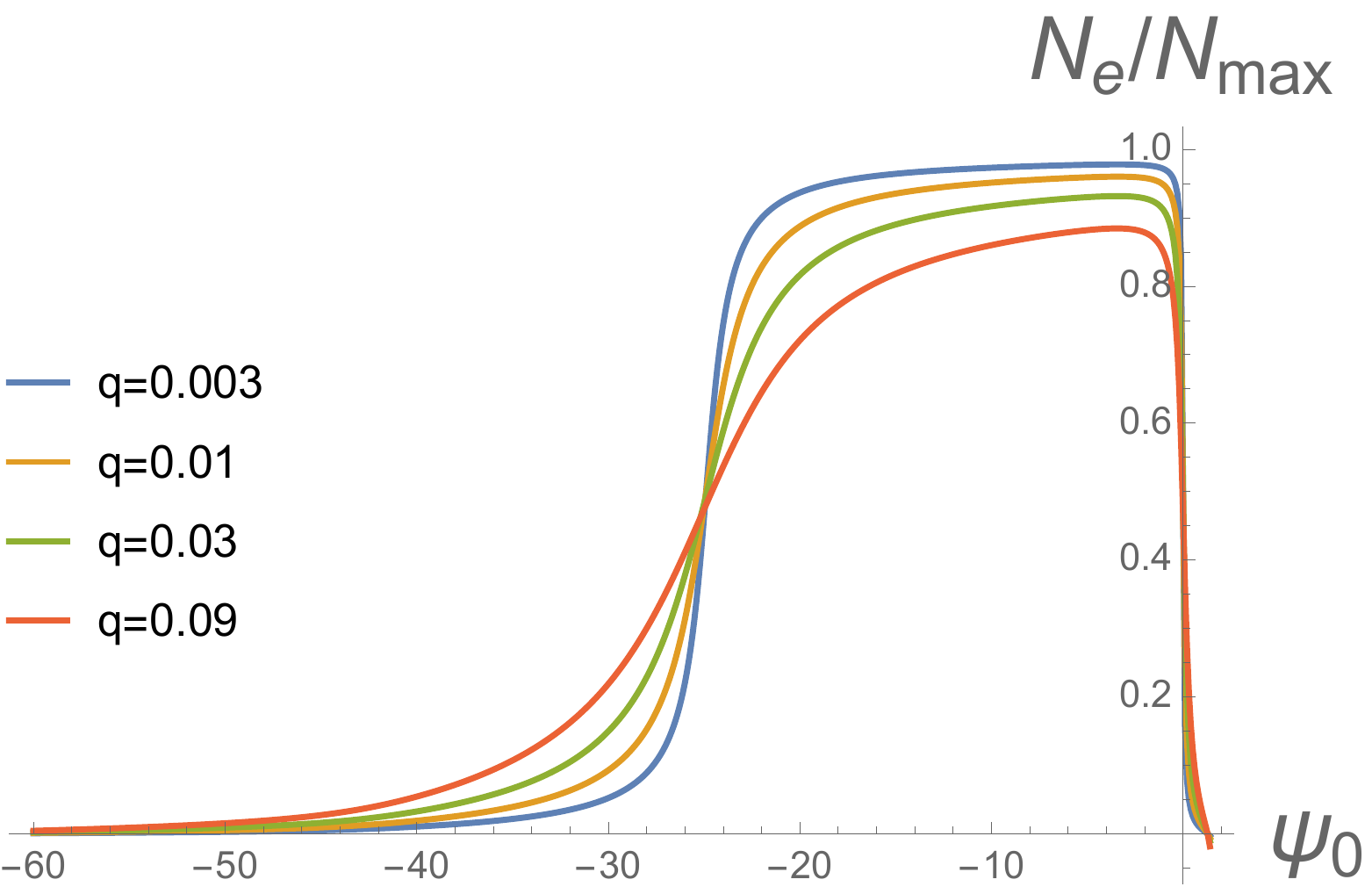}    
  \caption{The number of e-folds $N_e$ as a function of $\psi_0$ for several values of $q$ (left panel) and $N_e/N_{\rm max}$ (right panel).  
  }
  \label{fig:tend}
\end{figure}

We calculated numerically $N_e$ as a function of $\psi_0$ for several values of $q$.  The results are shown in Fig.\ref{fig:tend}.  We see that large values of $N_e \geq 50$ are reached only for $q\leq 0.03$.  For such values of $q$, $N_e \approx N_{\rm max}$ in almost the entire target range and $N_e$ is negligibly small outside of this range.  This is in full agreement with the above analysis.

We now comment on the validity of the approximation that we used to neglect the last term in Eq.~(\ref{psieq1}).  This term can be significant when $|\psi_*|\lesssim \sqrt{q} \approx 10/N_{\rm max}$, which corresponds to a narrow range of $\psi_0$ around $\psi_{\rm cr}$.  Using Eq.~(\ref{psi'}) we can estimate this range as
\beq
\Delta\psi_0 \sim \left| \frac{d\psi_*}{d\psi_0} (\psi_{\rm cr}) \right|^{-1} \frac{10}{N_{\rm max}}   \sim \frac{170}{N_{\rm max}} .
\label{Deltapsi0}
\eeq 
This is precisely the range where $N_e$ in Eq.~(\ref{Ne-psistar}) is significantly different from either $N_{\rm max}$ or 0.  For $\psi_0$ in this range, Eq.~(\ref{Ne-psistar}) for $N_e$ may not be very accurate, but we expect that $N_e$ interpolates between $N_{\rm max}$ and 0 over this range of $\psi_0$, so its qualitative behavior should be well represented by (\ref{Ne-psistar}).\footnote{The effect of a nonzero $q$ is to reduce $N_e$, so it can only make the transition region wider.  But an upper bound on the size of this region is set by Eq.~(\ref{Deltapsi0}); hence (\ref{Deltapsi0}) should be a good estimate for $\Delta\psi_0$.}

\subsection{Relation to earlier work}

The issue of overshooting in slow-roll inflation after tunneling has been discussed earlier by a number of authors \cite{Susskind,Dutta}.  Freivogel et al \cite{Susskind} assumed that the exit from tunneling is described by a potential with a linear slope, followed by a flat slow-roll region.  Dutta et al \cite{Dutta} considered power-law exit potentials, $U(\phi)\propto \phi^n$ with $n=1,2,3,4$.  The main conclusion following from this work is that if $\phi$ starts with some initial value $\phi_0$, it overshoots by an amount $\sim\phi_0$ into the slow-roll region.

In our context this would imply that tunnelings with initial values $\psi_0 \lesssim -1$ overshoot most, if not all, of the slow-roll region.  We find, however, that there is essentially no overshoot if the bubble nucleates with $\psi_0$ in a much wider range, $-25 \lesssim \psi_0\lesssim 0$.  The reason for this discrepancy may be that both Refs.~\cite{Susskind} and \cite{Dutta} assumed that $U_{\rm I}\ll U(\phi_0)$.  We find that, on the contrary, $U_{\rm I}$ is typically rather close to $U(\phi_0)$, $U(\phi_0) -U_{\rm I} \ll U_{\rm I}$.

\subsection{Probability of inflation}

 We can now estimate the probability for a randomly chosen inflection point $\phi_{\rm I}$ in the
  landscape to be a site of slow-roll inflation.  Inflation is possible only if the field tunnels out of
  the false vacuum to a point within the target range of $\phi_{\rm I}$.  The size of the target range is
  $\Delta\phi_{\rm target}\sim 25\Lambda^3$, while the distribution of the nucleation points has the
  width $\Delta\phi_0 \sim 0.4\Lambda$.  For $\Lambda\ll 0.1$, we have $\Delta\phi_{\rm
    target}\ll\Delta\phi_0$, and the probability for $\phi_0$ to be in the target range is \beq P_{\rm
    target} \sim \frac{\Delta\phi_{\rm target}}{\Delta\phi_0} \sim 60\Lambda^2 \ll 1.
\label{Ptarget}
\eeq

For most of inflection points, the tunneling then occurs outside the target range, so the inflaton overshoots and inflation does not happen.  On the other hand, for inflection points where $\phi_0$ is in the target range, inflation typically occurs over the entire slow-roll region, so $N_e \approx N_{\rm max}$.  Since $\phi_0$ is uncorrelated with with $N_{\rm max}$ for $N_{\rm max}\gg 1$, we expect that the probability distribution for $N_e$ is (almost) the same as that for $N_{\rm max}$.  The latter distribution has been calculated in Ref.~\cite{MVY},
\beq
P(N_{\rm max})dN_{\rm max} \approx {2\Lambda^4}\frac{dN_{\rm max}}{N_{\rm max}^3}.
\label{PMVY}
\eeq

Using Eqs.~(\ref{Ptarget}) and (\ref{PMVY}), we can now find the probability that a given inflection point supports inflation with $N_e >N$ e-folds,
\beq
P(>N)\sim P_{\rm target} \int_N^\infty P(N_{\rm max})dN_{\rm max} \sim \frac{60\Lambda^6}{N^2}.
\eeq
With $N\gtrsim 100$ and $\Lambda\ll 1$, this probability is rather small.  In our view, however, this is not an argument against the random Gaussian landscape model.  

Observational predictions of the model are based on landscape realizations with a large number of e-folds, sufficient to solve the flatness problem and to allow for structure formation.  All such realizations have $\phi_0$ in the target range.  Most of them have $N_e\approx N_{\rm max}$ and exhibit universal behavior.  If the observable scales lie within the slow-roll range, the spectral index $n_s$ is related to $N_{\rm max}$ by Eq.~(\ref{n_s for inflection}) and its distribution is given by Eq.~(\ref{Pns}).  The observed value of $n_s\approx 0.97$ is in the mid-range of this distribution, as discussed in \cite{MVY}.  Hence a random $1D$ Gaussian landscape is consistent with observations.

\subsection{Spatial curvature} 
\label{sec:distributions}

Since the interior geometry of the bubble is an open FRW universe, 
the curvature parameter is nonzero and may have an observable effect \cite{Susskind}.  At the present time this parameter is given by
\beq
 \Omega_K \simeq e^{-2 N_e} \lmk \frac{a_{\rm end} H_{\rm end}}{a_p H_p} \rmk^2, 
\eeq
where the subscripts ``end'' and ``p'' represent the values at the end of 
inflation and at present, respectively.  Hereafter, 
we assume instantaneous reheating and GUT scale inflation to give reference values. 
Then we have $\ln \Omega_K = - 2 (N_e - 62)$. 

The Planck collaboration puts an upper bound on $\Omega_K$ at $\Omega_K < 10^{-2} $, 
which requires 
\beq
 N_e \gtrsim 64. 
 \label{omega bound2}
\eeq
On the other hand, a detection of spatial curvature is probably possible only if $\Omega_K \gtrsim 10^{-4}$, or
\beq
 N_e \lesssim 67. 
 \label{omega bound3}
\eeq
Naively, we could use \eq{PMVY} to calculate the probability for $N_e$ to be within the range defined by these bounds.  This would give
\beq
P(64<N_e <67) = \frac{\int_{64}^{67} dN_e /N_e^3}{\int_{64}^\infty dN_e /N_e^3} \approx 2.5\% .
\eeq
We should note, however, that $N_e$ is correlated with $N_{\rm max}$, which is in turn related to the spectral index $n_s$ by Eq.~(\ref{n_s for inflection}).  
With the observed value of $n_s \approx 0.97$, we should have $N_{\rm max} \approx 120$.  Hence, in order to have observable curvature, $N_e$ has to be significantly smaller than $N_{max}$.  This is possible only if the tunneling point $\psi_0$ happens to be very close to the critical value $\psi_{\rm cr}$.

For $\abs{\psi_*} \ll 3 \pi / N_{\rm max}$, we can approximate \eq{Ne-psistar} as 
\beq
N_e \simeq \frac{N_{\rm max}}{2} \lkk 1 - \frac{2 N_{\rm max}}{3 \pi^2} \psi_* \rkk. 
\label{approxNe}
\eeq
The range of $\psi_0$ corresponding to the number of e-folds $N_e$ in the range of interest, $\Delta N_e \approx 3$, is then given by
\beq
\Delta\psi_0 = \frac{d\psi_0}{d\psi_*} \frac{d\psi_*}{dN_e} \Delta N_e,
\eeq
where the derivatives are evaluated at $\psi_*=0$.  Using Eqs.~(\ref{psi'}) and (\ref{approxNe}) we find
$\Delta\psi_0 \approx 500 N_{\rm max}^{-2}$ and
\beq
P(64<N_e <67) \approx \frac{\Delta\psi_0}{\psi_{\rm cr}} \approx 1.4\times 10^{-3}\left(\frac{N_{\rm max}} {120}\right)^{-2}.
\eeq
We thus see that the probability for the curvature to be smaller than the present upper bound but above the future detection limit is rather small.

\section{Conclusions}\label{sec:Conclusion}

We used numerical simulations to study bubble nucleation by quantum tunneling in random Gaussian potentials.  We were particularly interested in slow-roll inflation after the tunneling.  For a potential with a correlation length $\Lambda\ll 1$, this typically occurs near a flat inflection point $\phi_{\rm I}$, characterized by  $N_{\rm max} \gg 1$.  We sampled a large number of randomly generated potentials with flat inflection points and found that a substantial fraction of them (about a half) allow for tunneling from the false vacuum to the neighborhood of $\phi_{\rm I}$.  For each tunneling we found the initial value $\phi_0$ of the inflaton field in the bubble by solving the Euclidean field equation for the instanton.  The resulting distribution is peaked at $\phi_0\sim -0.1\Lambda$, where the inflection point is taken to be at $\phi=0$ and the positive direction of $\phi$ is taken to be towards the ``true'' vacuum.  The width of the distribution is $\sim 0.4\Lambda$ and is much larger than the size of the region where slow-roll inflation is possible, $\Delta\phi_{\rm sr} \sim \Lambda^3$.  This indicates that most of the tunnelings take the field $\phi$ outside of the slow-roll range.  

We developed a semi-analytic technique to study the evolution of $\phi$ after tunneling.  Our main conclusions can be stated as follows.  If the bubble nucleates with $\phi_0$ outside the slow-roll range, the field starts rolling fast (after a brief curvature-dominated period) and may overshoot part or all of the slow-roll region, or it may miss this region altogether.  We find that if $\phi_0$ is in the range 
\beq
\phi_{\rm cr}\lesssim \phi_0 \lesssim 0, 
\label{phicr}
\eeq
where $\phi_{\rm cr}\sim -25 \Lambda^3$, then the field slows down and undergoes a nearly maximal number of inflationary e-folds, $N_e\approx N_{\rm max}$.  On either side of the range (\ref{phicr}), $N_e$ drops towards zero within a distance $\Delta\phi_0 \sim \Lambda^3$.  The probability distribution for $N_e$ has the same power-law form as that for $N_{\rm max}$,
\beq
P(N_e)\propto N_e^{-3}. 
\label{PNe}
\eeq
The distribution for the spectral index $n_s$ is then given by Eq.~(\ref{Pns}) and is consistent with the observed value.

We also discussed the prospects for observational detection of nonzero spatial curvature $\Omega_K$, which is related to the number of e-folds $N_e$.  These prospects are not very good, because the observational lower bound on $N_e$ is pretty close to the upper bound that would make observational detection possible.  Using the distribution (\ref{PNe}), we found that the probability for a random observer in the multiverse to detect spatial curvature between these two bounds is $2.5\%$.  This estimate changes, however, if we take into account one more data point that is available to us: the measurement of the spectral index of density perturbations: $n_s\approx 0.97$.  In a small-field Gaussian landscape, $n_s$ is rigidly related to $N_{\rm max}$, and the observed value implies $N_{\rm max}\approx 120$.  On the other hand, the bound for future detectability requires that $N_e\lesssim 67$, which is significantly smaller than $N_{\rm max}$.  This situation is possible only if the tunneling point $\phi_0$ is very close to $\phi_{\rm cr}$ (within a range $\Delta\phi_0 \sim\Lambda^3$, where $N_e$ interpolates between $N_{\rm max}$ and 0).  Our estimate for the probability of this to happen is $\sim 10^{-3}$.  The bottom line is that the probability of detecting spatial curvature is pretty low if we live in a small-field random Gaussian landscape.

We finally comment on some limitations of our analysis.  The distribution for $\phi_0$ may have some additional factors, due to the probability measure of the multiverse.  These factors may depend on the tunneling transition rates between different vacua and on the correlation between the tunneling action and $\phi_0$.  This issue is entangled with the measure problem, which at present has no definitive solution and may require some new ideas to be resolved.

Another serious limitation is that we restricted our analysis to a one-dimensional landscape.   On the other hand, in multi-dimensional models parts of the landscape where slow-roll inflation is possible may be effectively one-dimensional, with only one of the fields having a nearly flat potential, due to an approximate shift symmetry.  In lieu of detailed information about the landscape, it may then be a reasonable approximation to consider such effective $1D$ potentials as samples of a random Gaussian field.  We note also that the methods we used here may have a wider applicability.  In Ref.~\cite{MVY} we indicated some directions in which these methods can be extended to multi-dimensional landscapes.

\section{Acknowledgement}
This work  is supported by the National Science Foundation under grant 1518742. M.Y. is supported by the JSPS Research Fellowships for Young Scientists. 
\appendix

\section{Validity of the interpolation method}\label{sec:singleValued}

In the main part of this paper, we generate the values of potential at equally spaced points $\phi_i$ ($i = 1,2,3, \dots n$) 
according to the joint distribution function $P( U_1, U_2, \ldots, U_n)$ obtained from correlation function in Eq.~\eqref{Correlation}. Here we defined $U_i=U(\phi_i)$.
We then smoothly interpolate these points to obtain $\tilde U(\phi)$. The value of $U(\phi)$ at any point is still a random variable. We can calculate the 
distribution of $U(\phi)$ with the prior knowledge of $\{U_1,\ldots, U_n \}$. To do so, we again calculate the joint distribution $P( U_1, U_2, \ldots, U_n, U(\phi))$ 
using the correlation function. We then calculate the conditional distribution of $U(\phi)$ using
\beq
\label{check}
 P(U(\phi)| U_1,\ldots,U_n)=\frac{P( U_1, U_2, \ldots, U_n, U(\phi))}{
P( U_1, U_2, \ldots, U_n)}. 
\eeq
To do this calculation let us define
\beq
 M_{ij} \equiv \la U(\phi_i) U(\phi_j) \ra~,
 \\
 M_{0 i } = M_{i 0} \equiv \la U(\phi_i) U(\phi) \ra~,
 \\
 M_{00} \equiv \la U(\phi) U(\phi) \ra, 
\eeq
where $i, j = 1,2,3, \dots, n$. 
Equation (\ref{check}) simplifies to 
\beq
  P(U(\phi)| U_1,\ldots,U_n)\propto\exp \lkk - \frac{1}{2 M^{-1}_{00}} \lmk U(\phi) + \sum_{j=1}^n M^{-1}_{0 j} 
 U(\phi_j) / M^{-1}_{00} \rmk \rkk. 
\eeq
Therefore, the mean value $\la U (\phi) \ra$ and width $\sigma$ are given by
\beq
 &&\la U (\phi)  \ra = - \sum_{j=1}^n M^{-1}_{0 j} 
 U(\phi_j) / M^{-1}_{00}, 
 \label{mean value}
 \\
 &&\sigma = \sqrt{M^{-1}_{00}}. 
 \label{1sigma}
\eeq

\begin{figure}[t] 
   \centering
   \includegraphics[width=2.7in]{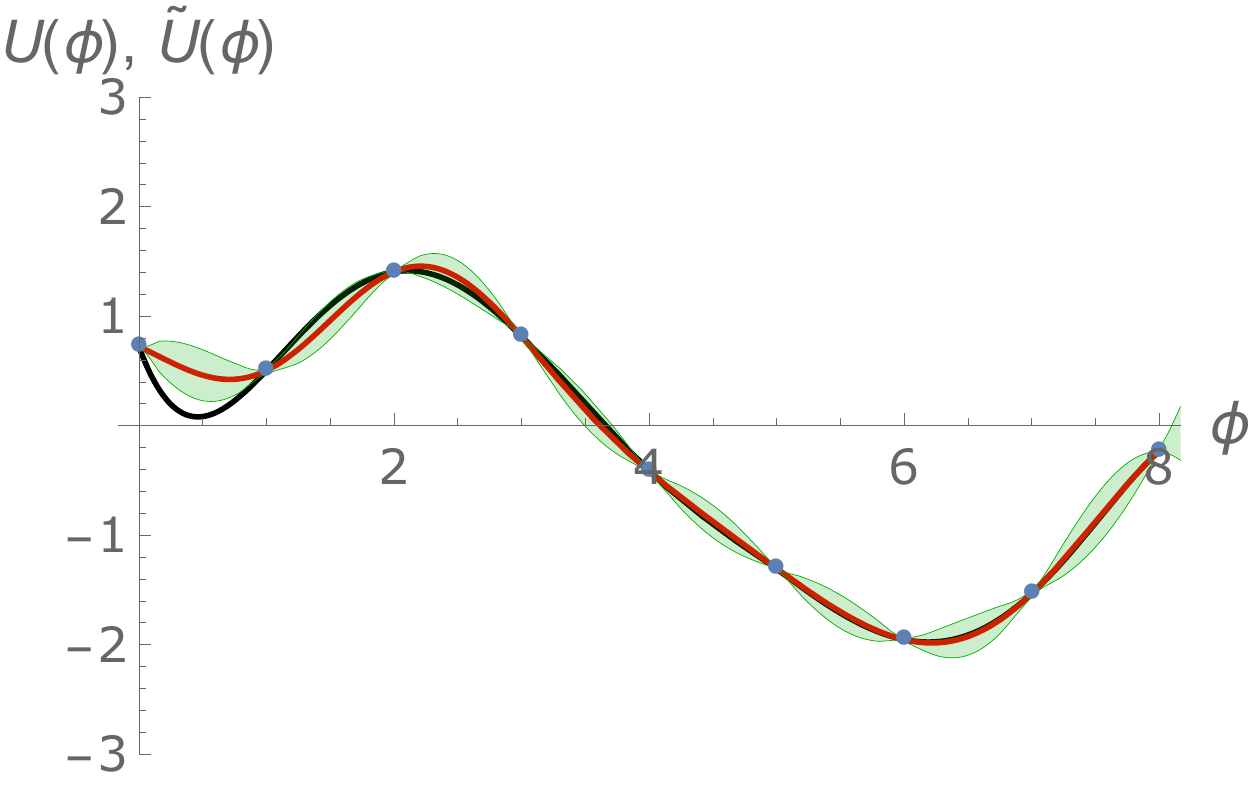} 
   \includegraphics[width=2.7in]{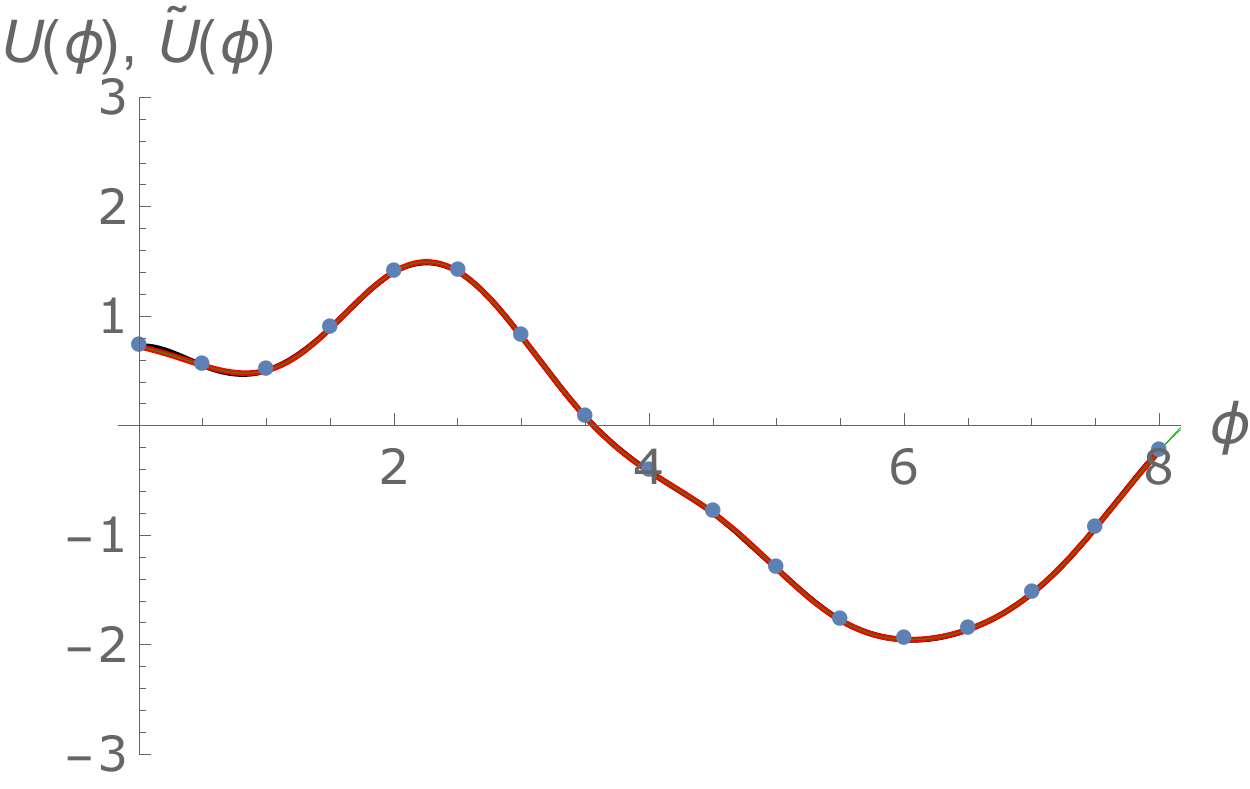}
   \includegraphics[width=2.7in]{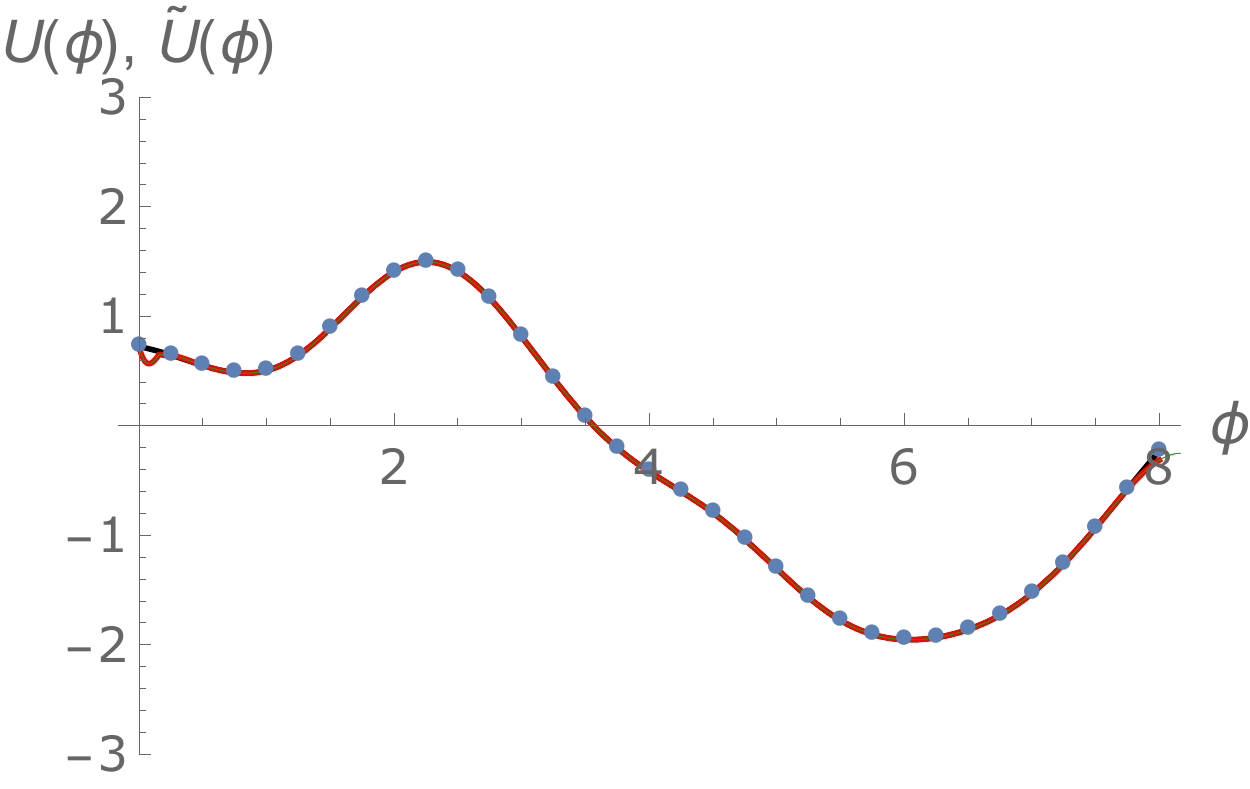} 
   \caption{
   Plots of realizations of potentials and their interpolating functions. 
   The dots represent realizations of potentials and the black lines are 
   the interpolating functions. 
   The number of points per unit correlation length is taken to be $1$ (upper left panel), 
   $2$ (upper right panel), and $4$ (lower panel). 
   The red lines are plotted by \eq{mean value} and the green regions are 
   its $2\sigma$ region given by \eq{1sigma}. 
   }
   \label{fig:check}
\end{figure}

We plot the interpolating function $\tilde{U} (\phi)$ and $\la U (\phi) \ra$ 
and $2 \sigma$ region according to \eq{1sigma} in Fig.~\ref{fig:check}. The $2\sigma$ region is so narrow
that we  can not see the difference between 
the interpolating function and mean value \eq{mean value} 
when the number of points per correlation length is larger than or equal to two. 
This result justifies the validity of interpolation method. 
In the main part of this paper, 
we take the number of points per correlation length as four 
and set $U(\phi) = \tilde{U} (\phi)$. 

\section{Cases of multiple instantons}\label{sec:uniqueness}
\begin{figure}[htbp] 
   \centering
   \includegraphics[width=2in]{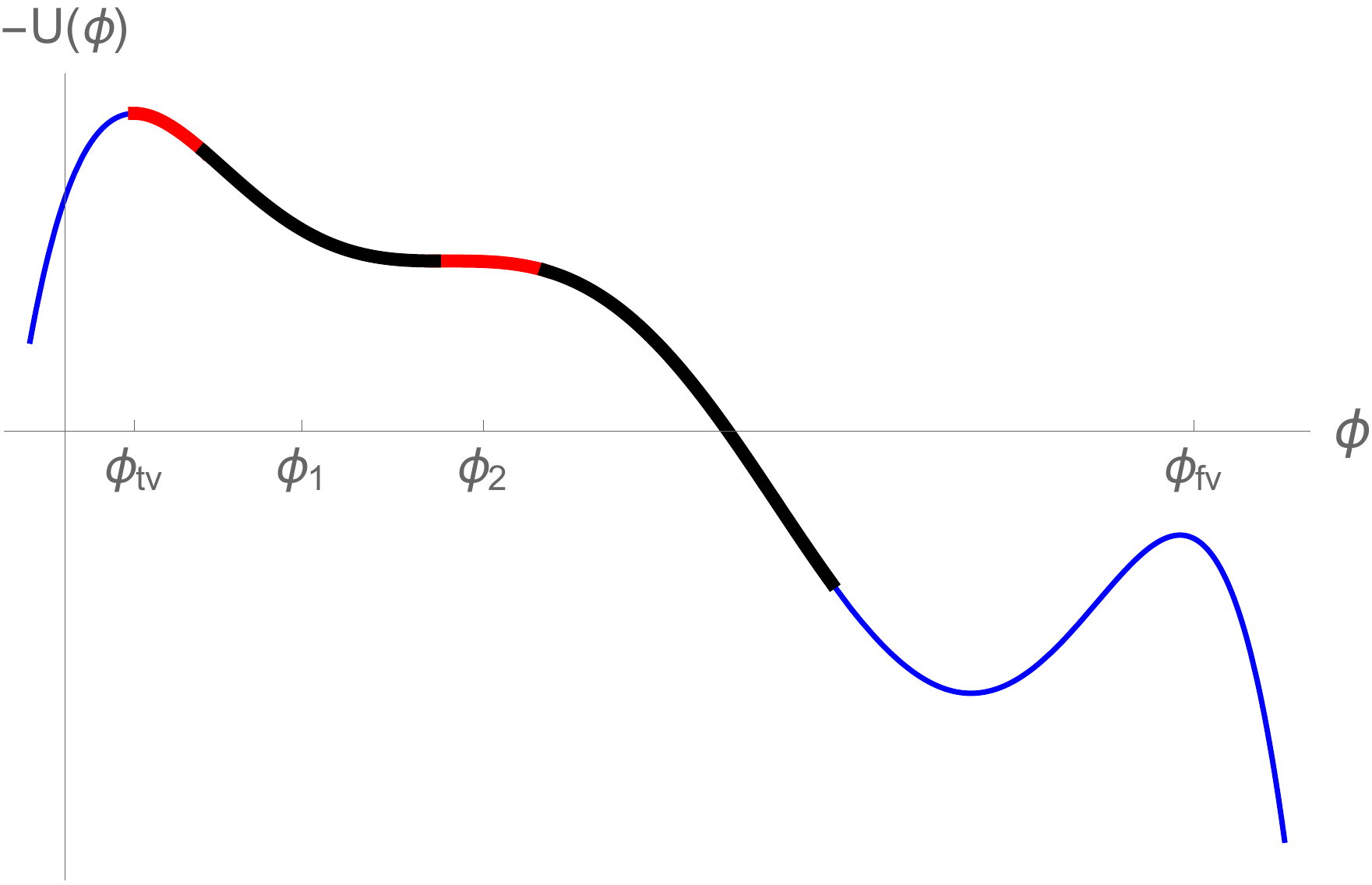} 
   \caption{An example of a potential which allows for more than one instanton solution.}
   \label{fig:TwoInst1}
\end{figure}

In this section we discuss the non-uniqueness of the instantons of one-dimensional potentials. We need to
solve Eq.~\eqref{quarticPot} with the boundary conditions $\phi'(0)=0$ and $\lim_{r \rightarrow \infty}
\phi(r)=\phi_{\rm fv}$. This is equivalent to the motion of a particle in a one-dimensional upside-down
potential, with $r$ playing the role of time. One can tune the value of the field at the center of the
instanton $\phi_0=\phi(0)$, so that starting from rest it ends up at rest at infinity. Using a simple
argument given by \cite{Coleman}, this is always possible. If $U(\phi_0)>U_{\rm fv}$ the field can never
reach the false vacuum. On the other hand, if $\phi_0$ is too close to the true vacuum, it stays there
for a while until the effect of the dissipating middle term is negligible. Without this dissipation, it
will pass the false vacuum, and hence in some place between these two extreme cases it must land on the
false vacuum. This suggests that the closer we start to the true vacuum, the less the dissipation
is. However, this is not always the case, as shown in Fig.\ref{fig:TwoInst1}.

Here we show in black the values of $\phi_0$ where the field never gets to the false vacuum and use red
for points where the field crosses the false vacuum in a finite time. Let's focus on $\phi_0=\phi_1$ and
$\phi_0=\phi_2$. The instanton starting at $\phi_1$ moves faster than the one starting at $\phi_2$ due to
the larger gradient at $\phi_1$. Hence, it loses more ``energy'' to the dissipative middle term. The
difference between the energy losses can be more than $U(\phi_1)-U(\phi_2)$, and hence $\phi_0=\phi_2$
overshoots while $\phi_0=\phi_1$ undershoots. At the boundary of red and black regions, there are points
that satisfy the boundary conditions. For example in this case there are three such solutions. This shows
that new instantons are created in pairs and (except for measure zero set) there is always an odd number
of instantons. Let's mark the instantons from left to right by 1, 2, \ldots. Although we do not have a
proof, our numerical solutions for thousands of instantons suggest that the instantons with even numbers
have higher actions and are always sub-dominant. We do not know if these instantons have the right number
of negative modes and whether they contribute to tunneling at all.

\section{Possibility of favorable tunneling to far away minima}\label{sec:tunnelingFar}
\begin{figure}[htbp]
  \centering
  \includegraphics[width=3in]{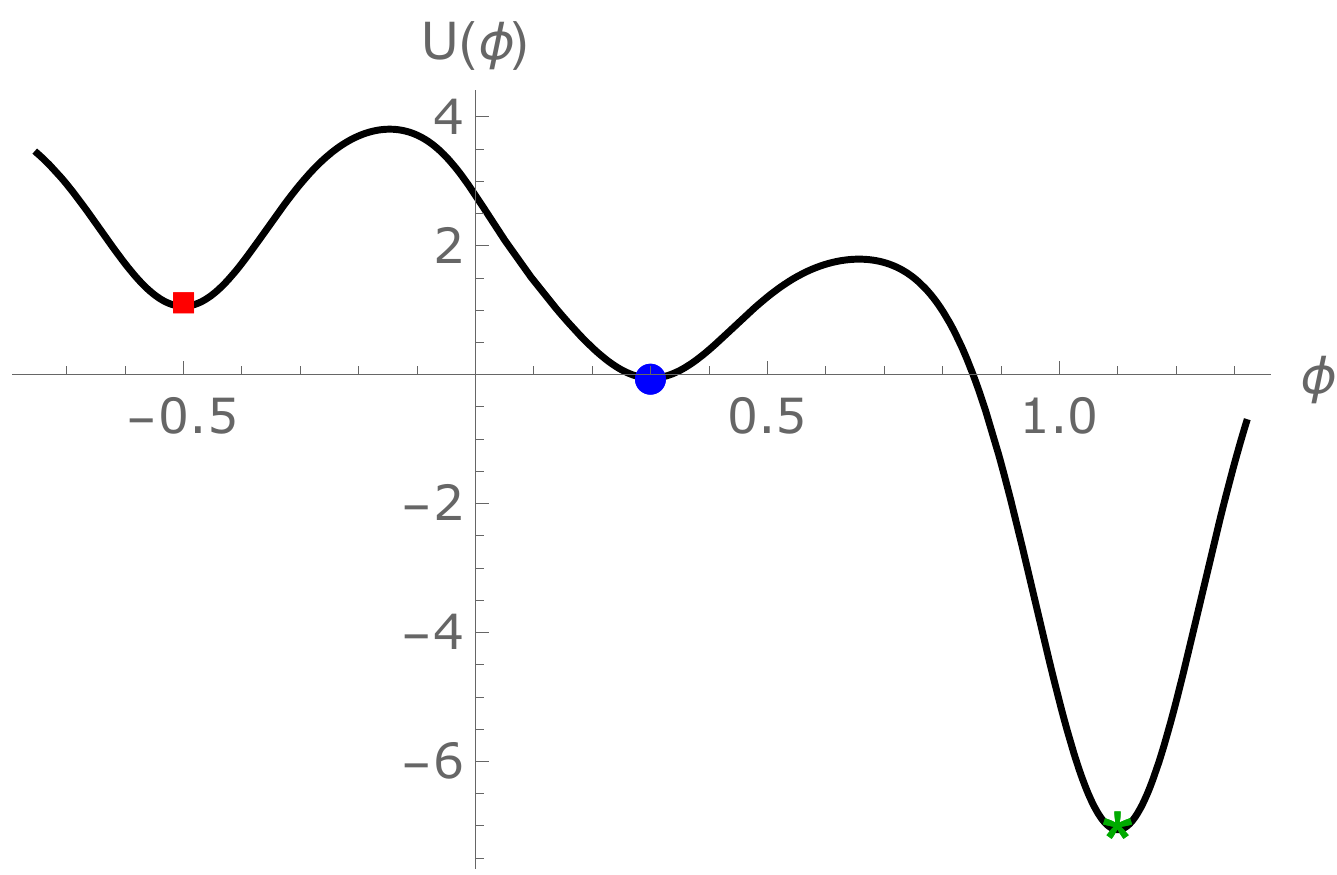}
  \caption{An example of a potential in which decay to the further away minimum is dominant. The metastable
    vacuum on the left (red square) can decay both to the middle (blue circle) and the right
    (green star) minimum. The action to the former is around 297 and the latter around 62. Hence, the
    left vacuum dominantly decays directly to the vacuum on the right.}
  \label{fig:FarMinima}
\end{figure}

In this appendix we present another new and counter-intuitive feature of tunneling in theories with one
field.\footnote{Similar phenomena can happen in multi-field case.} We show that it is possible for a
metastable vacuum to decay dominantly to minima which are further away rather than the closer ones. It
is not difficult to find cases with  two vacua, one on the left and one on the right of a
given metastable vacuum, where tunneling to the further away minimum is dominant granted that the barrier
separating them is lower. What is non-intuitive is the existence of cases where both  minima are on
the same  side of a false vacuum and the decay to the close minimum is suppressed. One such  example is
shown in  Fig.\ref{fig:FarMinima}.
If there was not a minimum in the middle of the right and left minima, there would always be an instanton
that carries the decay to the vacuum on the right. However, if the minimum in the middle (blue circle
minimum) is not very deep, it does not effect this instanton much. According to Coleman, there is always
an instanton that causes decay to the blue circle minimum. However, this vacuum is nearly degenerate with
the one on the left, so we expect it to have a large action and hence to be subdominant.

\end{document}